\documentclass[12pt,preprint]{aastex}

\begin{document}

\title{Determination of Nucleosynthetic Yields of Supernovae
and Very Massive Stars from Abundances in Metal-Poor Stars}
\author{Y.-Z. Qian\altaffilmark{1} and G. J. Wasserburg\altaffilmark{2}}
\altaffiltext{1}{School of Physics and Astronomy, University of
Minnesota, Minneapolis, MN 55455; qian@physics.umn.edu.}
\altaffiltext{2}{The Lunatic Asylum, Division of Geological and
Planetary Sciences, California Institute of Technology, Pasadena,
CA 91125.}

\begin{abstract}
We determine the yields of the elements from Na to Ni for Type II
supernovae (SNe II) and the yield patterns of the same elements
for Type Ia supernovae (SNe Ia) and very massive 
($\gtrsim 100\,M_\odot$) stars (VMS) using a phenomenological model 
of stellar nucleosynthesis and the data on a number of stars with
$-4\lesssim{\rm [Fe/H]}\lesssim -3$, a single star with 
[Fe/H]~$=-2.04$, and the sun. We consider that there are two distinct
kinds of SNe II: the high-frequency SNe II($H$) and the low-frequency 
SNe II($L$). We also consider that VMS were the dominant 
first-generation stars formed from big bang debris.
The yield patterns of Na to Ni for SNe II($H$), II($L$), and Ia and
VMS appear to be well defined. It is found that SNe II($H$) produce
almost none of these elements, that SNe II($L$) can account for the
entire solar inventory of Na, Mg, Si, Ca, Ti, and V, and that
compared with SNe II($L$), VMS underproduce Na, Al, V, Cr, and Mn,
overproduce Co, but otherwise have an almost identical yield pattern. 
A comparison is made
between the yield patterns determined here from the observational
data and those from ab initio models of nucleosynthesis in SNe II
and VMS.

The evolution of the other elements relative to Fe is shown to involve
three distinct stages. The earliest stage is in the domain of
[Fe/H]~$< -3$ and is governed by VMS activities with some small
contributions from SNe II, all of which are dispersed in a
dilution mass of $M_{\rm dil}^{\rm VMS}\sim 10^6$--$10^7\,M_{\odot}$. 
The beginning of the second stage is marked by
the cessation of VMS activities and the onset of major formation of
normal stars (with masses of $\sim 1$--$60\,M_{\odot}$) at 
[Fe/H]~$\approx -3$. The dilution mass for SN II contributions
also drops sharply to 
$M_{\rm dil}^{\rm SNII}\approx 3\times 10^4\,M_{\odot}$.
The subsequent quasi-continuous chemical evolution until
[Fe/H]~$\sim -1$ is governed by SNe II($H$), which produce mainly
the heavy $r$-process elements above Ba, and SNe II($L$), which
produce essentially all the other elements. The third stage starts
with the onset of SN Ia contributions to mainly the Fe group
elements at [Fe/H]~$\sim -1$. The domain of [Fe/H]~$> -1$ is then
governed by contributions from SNe II($H$), II($L$), and Ia and 
other low mass stars. It is shown
that the abundances of non-neutron-capture elements
in stars with [Fe/H]~$\leq 0$ and those of $r$-process elements
in stars with [Fe/H]~$<-1$ can be well
represented by the sum of the distinct components in the 
phenomenological model. The proposed evolutionary sequence is 
directly related to
the problems of early aggregation and dispersion of baryonic
matter and to the onset of formation and chemical evolution of
galaxies. It is argued that the prompt inventory governed by VMS
contributions should represent the typical composition of
dispersed baryonic matter in the universe. 
\end{abstract}


\section{Introduction}

In this paper we discuss the nucleosynthetic yields of supernovae 
(SNe) and very
massive stars (VMS, with masses of $\gtrsim 100\,M_\odot$) 
based on the observed abundances in stars with
$-4\lesssim{\rm [Fe/H]}<-1$. It was considered in our previous
works [Wasserburg \& Qian 2000a (WQ),b; Qian \& Wasserburg
2001a,c (QW)] that VMS were formed from big bang debris and provided
an initial or prompt ($P$) inventory of Fe and associated elements
up to [Fe/H]~$\approx -3$. It was assumed that normal stars (with
masses of $\sim 1$--$60\,M_\odot$) could only be formed at
[Fe/H]~$\gtrsim -3$. Subsequent evolution of normal stars led to
Type II SNe (SNe II) and at much later times (corresponding to
[Fe/H]~$\gtrsim -1$) also to Type Ia SNe (SNe Ia), all of which
provided further chemical enrichment of the interstellar medium
(ISM) beyond the $P$-inventory. The special status of
[Fe/H]~$\approx -3$ in chemical evolution of the ISM was
identified based on a sharp increase in the abundances of heavy
$r$-process elements (Ba and above) at this metallicity
(WQ). This is shown in Figure \ref{figba} for
Ba. The sharp increase was attributed to the rapid occurrence of
high-frequency SNe II($H$) that produce mainly the heavy
$r$-process elements but no Fe. The Fe enrichment of the ISM at
$-3<{\rm [Fe/H]}< -1$ was attributed to the low-frequency SNe
II($L$) and that at [Fe/H]~$\gtrsim -1$ to both SNe II($L$) and Ia.
In addition to contributing $\approx 1/3$ of the solar Fe inventory,
SNe II($L$) are mainly responsible for the light $r$-process
elements (Ba and below). QW have derived the $P$-inventory and SN
II($H$) and II($L$) yields of $r$-process elements from the
observed abundances in two stars with [Fe/H]~$\approx -3$ and the
solar $r$-process abundances. They have shown that the abundances
calculated from the three-component model including the
$P$-inventory and the contributions from SNe II($H$) and II($L$)
were in good agreement with several independent data sets for a
large number of $r$-process elements over the wide range of
$-3\lesssim{\rm [Fe/H]}< -1$. The same agreement was also
achieved for Sr, Y, Zr, and Ba when the standard solar $r$-process
abundances of these elements (Arlandini et al. 1999) were
increased by factors of $\approx 2$--6. In general, the model of
QW gave a satisfactory description of the chemical evolution of
all the $r$-process elements over $-3\lesssim{\rm [Fe/H]}< -1$,
which corresponds to a period of $\approx 3\times 10^9$~yr in
the early Galactic history.

Here we extend the model of QW to the elements from Na to Ni and
derive the relevant yields of SNe II($H$) and II($L$) and the
yield pattern of SNe Ia (\S\ref{secysn}). We also infer the yield
pattern of VMS from the observed abundances in stars with
$-4\lesssim{\rm [Fe/H]}\lesssim -3$ (\S\ref{secyvms}). 
It will be shown
that SNe II($H$) produce almost none of the elements from Na to
Ni, that SNe II($L$) can account for the entire solar inventory of
Na, Mg, Si, Ca, Ti, and V, and that compared with SNe II($L$), VMS
underproduce Na, Al, V, Cr, and Mn, overproduce Co, but otherwise
have an almost identical yield pattern. We re-interpret the
metallicity of [Fe/H]~$\approx -3$ as corresponding to the
cessation of VMS activities and the onset of major formation of
normal stars. It will be shown that VMS dominated the production of
Na to Ni at [Fe/H]~$\lesssim -3$ and that explosions of these
objects could induce mixing within $\sim 10^6$--$10^7\,M_\odot$ 
of ISM, to be
compared with $\approx 3\times 10^4\,M_\odot$ for SNe II. A model
for general chemical evolution of the ISM starting from a zero
metallicity state will be presented (\S\ref{secevol}). The wide
range in yields calculated from nucleosynthetic models 
for SNe II and VMS will be 
discussed in light of the rather constant and 
metallicity-independent yield
patterns as determined here
from the observed abundances in metal-poor stars
(\S\ref{secdis}). The cosmological implications of the results 
presented here will also be discussed (\S\ref{secdis}).

\section{Three-Component Model and the $P$-Inventory and SN Yields}
\label{secysn}

We consider a homogeneous mass of gas and assume
that stars formed from this gas have the composition of the gas.
The chemical evolution of an element E in this gas is specified by
the initial number of E atoms per H atom in the gas
and the number of E atoms per H atom added from various SNe II to the gas.
As shown by QW, if each distinct kind of SN II has a fixed relative yield 
pattern, then the inventory of E atoms may be simply described
by the effective numbers of SNe II that contributed to the gas.
The effective numbers are calculated by assuming constant yields and a
standard dilution mass for SNe II. 
The results of astration, fragmentation, and 
merging of different gas masses, as well as the effects of variable 
SN II yields (but with fixed yield patterns), are subsumed in the 
representation of the 
effective numbers of contributing SNe II. By using this approach,
it was shown that the abundances of $r$-process and related
elements (Sr and above) in stars with $-3\lesssim{\rm [Fe/H]}<-1$
can be described by a three-component model including the
$P$-inventory and the contributions from SNe II($H$) and II($L$)
(QW). For a star formed from an ISM with a number $n_H$ of
contributing SNe II($H$) and a number $n_L$ of contributing SNe
II($L$), the abundance of an element E in the star is given
by
\begin{equation}
\label{eps}
10^{\log\epsilon({\rm E})}=10^{\log\epsilon_P({\rm E})}+n_H\times
10^{\log\epsilon_H({\rm E})}+n_L\times 10^{\log\epsilon_L({\rm E})},
\end{equation}
where $\log\epsilon_P({\rm E})$, $\log\epsilon_H({\rm E})$, and
$\log\epsilon_L({\rm E})$ represent the $P$-inventory and the SN II($H$)
and II($L$) yields of E, respectively. Here and below, the
standard spectroscopic notation $\log\epsilon({\rm E})\equiv\log\epsilon
({\rm E/H})+12$ is used. The
parameter $\log\epsilon_H({\rm E})$ or $\log\epsilon_L({\rm E})$
corresponds to the abundance of E in the ISM resulting
from a single SN II($H$) or II($L$) for a standard dilution mass
($M_{\rm dil}^{\rm SN II}\approx 3\times 10^4\,M_\odot$) of baryonic 
matter that is free of ``metals'' and dominantly composed of H. 
This dilution mass is fixed by the frequency of SNe II($H$) as required
for replenishment of fresh radioactive $^{182}$Hf in the ISM (e.g.,
Wasserburg, Busso, \& Galllino 1996; QW)
and is consistent with the typical total amount of ISM swept up by an
SN II remnant (e.g., Thornton et al. 1998).

As the heavy $r$-process elements above Ba,
e.g., Eu, are exclusively produced by SNe II($H$), the number $n_H$ for
the star can be obtained from its observed $\log\epsilon({\rm Eu})$:
\begin{equation}
\label{nh}
n_H=10^{\log\epsilon({\rm Eu})-\log\epsilon_H({\rm Eu})}=
10^{\log\epsilon({\rm Eu})+2.48},
\end{equation}
where $\log\epsilon_H({\rm Eu})=-2.48$ (QW) is used. Likewise,
as additions of
Fe at $-3<{\rm [Fe/H]}<-1$ beyond the $P$-inventory are made only by
SNe II($L$), the number $n_L$ for the star can be obtained from its
observed $\log\epsilon({\rm Fe})$:
\begin{equation}
\label{nle}
n_L=\left[10^{\log\epsilon({\rm Fe})}-10^{\log\epsilon_P({\rm Fe})}
\right]/10^{\log\epsilon_L({\rm Fe})},
\end{equation}
where $\log\epsilon_P({\rm Fe})=4.51$ and $\log\epsilon_L({\rm
Fe})=5.03$ (QW). By using ${\rm [Fe/H]}\equiv
\log\epsilon({\rm Fe})-\log\epsilon_\odot({\rm Fe})$ with
$\log\epsilon_\odot({\rm Fe})=7.51$ (Anders \& Grevesse 1989),
equation (\ref{nle}) can be rewritten as
\begin{equation}
\label{nl}
n_L=\left(10^{\rm [Fe/H]}-10^{{\rm [Fe/H]}_P}\right)/
10^{{\rm [Fe/H]}_L}=10^{{\rm [Fe/H]}+2.48}-10^{-0.52},
\end{equation}
where [Fe/H]$_P=-3$ and [Fe/H]$_L=-2.48$ are used. The parameters
$\log\epsilon_P$, $\log\epsilon_H$, and $\log\epsilon_L$ for the
pertinent elements were obtained from the data on two stars with
[Fe/H]~$\approx -3$ and the solar $r$-process abundances (QW).
It was shown that the abundances calculated from the
above model for a large number of $r$-process elements were in
good agreement with several independent data sets over the
wide range of $-3\lesssim{\rm [Fe/H]}<-1$.
The same agreement was also achieved for
Sr, Y, Zr, and Ba when the standard solar $r$-process abundances
of these elements (Arlandini et al. 1999) were increased by
factors of $\approx 2$--6.

\subsection{Estimates of the $P$-Inventory and SN II($H$) Yields}

We now extend the above three-component model to the elements
below Sr. By the assumptions of the model, a fixed amount of ISM
dilutes the ejecta from an SN II at [Fe/H]~$\gtrsim -3$. 
It can be seen
from equation (\ref{nl}) that for the standard dilution mass,
stars with [Fe/H]~$\approx -3$ have $n_L\approx 0$. So the
abundances in these stars only represent mixtures of the
$P$-inventory and SN II($H$) contributions:
\begin{equation}
\label{ph}
10^{\log\epsilon({\rm E})}=10^{\log\epsilon_P({\rm E})}+n_H\times
10^{\log\epsilon_H({\rm E})}.
\end{equation}
Thus, with the value of $n_H$ obtained from equation (\ref{nh}), the
parameters $\log\epsilon_P({\rm E})$ and $\log\epsilon_H({\rm E})$
can be determined from the data on any two stars with [Fe/H]~$\approx
-3$ but different $n_H$ values.

Figure \ref{figp}a shows the observed $\log\epsilon$ values of C
to Ge (Westin et al. 2000; McWilliam et al. 1995) for HD 122563
([Fe/H]~$=-2.74$; open squares), HD 115444 ([Fe/H]~$=-2.99$;
filled circles), and CS 22892-052 ([Fe/H]~$=-3.03$; asterisks)
with $n_H\approx 1$, 7, and 36, respectively. The typical
uncertainties in the data shown in Figure \ref{figp}a are $\approx
0.1$~ dex. It can be seen that the abundances of
the elements above C are essentially the same for all three stars.
In fact, when a uniform shift of $-0.25$~dex is applied to the
data on HD 122563 so that all three sets of data correspond to
essentially the same [Fe/H], significant differences are only
found for C between the three stars and for Ge between HD 122563
and HD 115444. The differences for all the other elements are
within the observational uncertainties. As these three stars have
very different $n_H$ values, we conclude that SNe II($H$) produce
very little of the elements from N to Zn. The abundances of these
elements in the three stars are then completely dominated by the
$P$-inventory. We take the observed $\log\epsilon$ values of N to
Zn for HD 115444 to be the corresponding $\log\epsilon_P$ values,
thus defining the abundances in the $P$-inventory.

An upper limit on $\log\epsilon_H$ for N to Zn can be
estimated from
\begin{equation}
\label{yh}
10^{\log\epsilon_P({\rm E})+\sigma({\rm E})}\gtrsim
10^{\log\epsilon_P({\rm E})}+n_H^{\rm HD 115444}
\times 10^{\log\epsilon_H({\rm E})},
\end{equation}
where $\sigma({\rm E})\approx 0.1$~dex is the uncertainty in the
observed $\log\epsilon({\rm E})$ value for HD 115444 and $n_H^{\rm
HD 115444}\approx 7$. Equation (\ref{yh}) gives
\begin{equation}
\label{yhu}
\log\epsilon_H({\rm E})\lesssim\log\epsilon_P({\rm E})-1.43
\end{equation}
for N to Zn. Thus the contributions from an SN II($H$) with the
standard dilution mass are small compared with the existing 
$P$-inventory for all the elements between C and Ge.

The $\log\epsilon_P$ values and the upper limits on
$\log\epsilon_H$ estimated in equation (\ref{yhu}) are given in
Table \ref{tabyphl}. It was argued that the extremely high value
of $n_H\approx 36$ corresponding to the extremely high $r$-process 
abundances observed in CS 22892-052 is due to contamination of
the surface of this star by the ejecta from the SN II($H$)
explosion of a previous massive binary companion (Qian \&
Wasserburg 2001b). The agreement between the observed
$\log\epsilon$ values of all the elements from Na to Ni for CS
22892-052 and the corresponding $\log\epsilon_P$ values in Table
\ref{tabyphl} further supports the argument that the SN II($H$)
yields of these elements are very low. On the other hand, the
$\log\epsilon({\rm C})$ value of CS 22892-052 is higher than those
of HD 115444 and HD 122563 by $\approx 1$~dex 
(see Figure \ref{figp}a). This suggests that
a significant amount of C may be produced in SNe II($H$) or that C
was produced in CS 22892-052 during its evolution. The elements C,
N, O, Cu, Zn, and Ge will not be discussed further [the evolution
of O relative to Fe has been discussed in detail in a previous
paper (Qian \& Wasserburg 2001a)]. Below we will focus on the
elements from Na to Ni as there are high-quality data on these
elements.

Figure \ref{figp}b compares the $P$-inventory (see Table
\ref{tabyphl}) as determined above for [Fe/H]~$=-3$ with
the observational data at $ -4\lesssim [{\rm Fe/H}] < -3$. We use
the results of Norris, Ryan, \& Beers (2001) on CD $-24^\circ
17504$ ([Fe/H]~$=-3.37$; squares), CD $-38^\circ 245$
([Fe/H]~$=-3.98$; diamonds), CS 22172-002 ([Fe/H]~$=-3.61$;
triangles), CS 22885-096 ([Fe/H]~$=-3.66$; asterisks), and CS
22949-037 ([Fe/H]~$=-3.79$; pluses). We have shifted the
$\log\epsilon$ values shown for these five stars from the observed
values to pass through $\log\epsilon({\rm Fe})=\log\epsilon_P({\rm
Fe})=4.51$ so that they are compared at the same Fe abundance. It
can be seen that the abundance patterns in these stars follow the
$P$-inventory rather closely. Of these stars, CS 22949-037
exhibits the largest deviation from the $P$-inventory as it is
overabundant in Mg and Si. Data on this star from McWilliam et al.
(1995) also show that it is overabundant in Na. However, the
abundance pattern of all the other elements and the ratio Si/Mg
for this star are essentially the same as for the $P$-inventory.
We thus consider that the $P$-inventory determined by the data at
[Fe/H]~$\approx -3$ is also an excellent representation of the
data at $-4\lesssim [{\rm Fe/H}]< -3$. We infer that VMS must
therefore have a fairly uniform nucleosynthetic mechanism if they
were responsible for the $P$-inventory. In \S\ref{secyvms} we will
show that the $P$-inventory must also contain some contributions
from SNe II($H$) and II($L$).

\subsection{Estimates of SN II($L$) Yields and SN Ia Yield Pattern}

We now derive the SN II($L$) yields of Na to Ni
based on the three-component model.
By using the $\log\epsilon_P$ and $\log\epsilon_H$ values
in Table \ref{tabyphl} together with equation (\ref{eps}),
the $\log\epsilon_L$ values of these elements can be obtained from
the observed $\log\epsilon$ values for HD 178443 with $n_L=2.45$
([Fe/H]~$=-2.04$; McWilliam et al. 1995). The results
are given in Table \ref{tabyphl}. It can be seen that typically
$\log\epsilon_L({\rm E})$ exceeds $\log\epsilon_H({\rm E})$ by
$\approx 2$ dex. Thus, although SNe II($H$) are 10 times more
frequent than SNe II($L$) (e.g., QW), numerically the SN II($H$)
contributions to the elements from Na to Ni can be ignored in
general.

Using the SN II($H$) and II($L$) yields, we give in Table
\ref{tabyphl} the fraction $\alpha_{\odot,{\rm II}}({\rm E})$ of
the solar inventory of an element E contributed by SNe II:
\begin{equation}
\label{aii}
\alpha_{\odot,{\rm II}}({\rm E})\equiv[n_H^\odot
\times 10^{\log\epsilon_H({\rm E})}+n_L^\odot
\times 10^{\log\epsilon_L({\rm E})}]/
10^{\log\epsilon_\odot({\rm E})},
\end{equation}
where $n_H^\odot=10^3$ and $n_L^\odot=10^2$ (QW)
are the numbers of SNe
II($H$) and II($L$), respectively, that had occurred in a standard
dilution mass of ISM prior to solar system formation (SSF). The
contributions from the $P$-inventory to the solar inventory are
negligible. Table \ref{tabyphl} shows that SNe II($L$) can
essentially account for the entire solar inventory of Na, Mg, Si,
Ca, Ti, and V [$\alpha_{\odot,{\rm II}}({\rm E})\approx 1$ for
these elements as $\log\epsilon_L({\rm E}) + \log n_L^\odot\approx
\log\epsilon_{\odot}({\rm E})$].  By contrast, Table \ref{tabyphl}
shows that the $\alpha_{\odot,{\rm II}}$ values for Al, Sc, Cr,
Mn, Fe, Co, and Ni only range from 0.13 to 0.47. In view of the
uncertainties in the data used to derive the $\log\epsilon_L$
values, we now choose to assign the entire solar inventory of Na,
Mg, Si, Ca, Ti, and V to SNe II($L$) [i.e., we take the 
``corrected'' values of
$\log\epsilon_L^{\rm corr}({\rm E})=\log\epsilon_{\odot}({\rm E})
-\log n_L^\odot$ and $\alpha_{\odot,{\rm II}}^{\rm corr}({\rm E})
=1$ for these elements]. For Al, Sc, Cr, Mn, Fe, Co, and Ni that
have $\alpha_{\odot,{\rm II}}$ values significantly below unity,
we will use the original calculated values [i.e.,
$\log\epsilon_L^{\rm corr}({\rm E})=\log\epsilon_L({\rm E})$
and $\alpha_{\odot,{\rm II}}^{\rm corr}({\rm E})=
\alpha_{\odot,{\rm II}}({\rm E})$ for these elements].
The SN II($L$) yields represented by the 
$\log\epsilon_L^{\rm corr}$ values in Table \ref{tabyphl}
are shown in Figure \ref{figyl} along 
with the solar abundance pattern that is translated to pass
through $\log\epsilon_L^{\rm corr}({\rm Mg})$.

The majority of the solar inventory of the
so-called Fe group elements (Cr, Mn, Fe, Co and Ni) may be
reasonably attributed to SNe Ia (e.g. Timmes, Woosley, \& Weaver
1995). In this case, the part of the solar inventory of an element E
in the Fe group contributed by SNe Ia,
$\log\epsilon_{\odot,{\rm Ia}}({\rm E})$, can be estimated as
\begin{equation}
\label{yia}
\log\epsilon_{\odot,{\rm Ia}}({\rm E})=
\log[1-\alpha_{\odot,{\rm II}}({\rm E})] +\log\epsilon_\odot({\rm E}).
\end{equation}
The $\log\epsilon_{\odot,{\rm Ia}}$ values for
Cr, Mn, Fe, Co and Ni estimated in
equation (\ref{yia}) represent the yield pattern of these elements
for SNe Ia and are given in Table \ref{tabyphl}.

\section{The $P$-Inventory and VMS Yield Pattern}
\label{secyvms}

We have established the $P$-inventory and SN II($H$) and II($L$)
yields of Na to Ni in \S\ref{secysn}. We now focus on the
$P$-inventory. This inventory was considered to be the result of
the integrated production by VMS over the period prior to the
achievement of a
metallicity of [Fe/H]~$\approx -3$ in the ISM (WQ).
Figure \ref{figp}b shows that the $P$-inventory is also an
excellent representation of the abundance patterns at
$-4\lesssim{\rm [Fe/H]}< -3$. This suggests that the yields of VMS
must follow a rather regular pattern. We will seek to establish
the yield pattern of VMS from the data at $-4\lesssim{\rm [Fe/H]}
\lesssim -3$. It is important to recognize that very old stars with
$-4\lesssim{\rm [Fe/H]}< -3$ must have low masses of $\sim
1\,M_\odot$ to have survived until the present time. This requires
that there be some formation of stars in the normal mass range of
$\sim 1$--$60\,M_\odot$ in addition to the formation of VMS with
masses of $\gtrsim 100\,M_\odot$ considerably before a metallicity
of [Fe/H]~$\approx -3$ was reached in the ISM. As normal stars
with masses of $\sim 10$--$60\,M_\odot$ are SN II progenitors,
we consider that SNe II($H$) and II($L$) must have provided
chemical enrichment of the ISM along with VMS at $-4\lesssim{\rm
[Fe/H]}< -3$. Thus, the $P$-inventory at [Fe/H]~$=-3$ and the
abundances in stars with $-4\lesssim{\rm [Fe/H]}< -3$ should be
the sum of contributions from VMS and SNe II($H$) and II($L$). The
yield pattern of VMS must then be obtained by subtracting the SN
II($H$) and II($L$) contributions. This can be accomplished by
using the observed Ba abundances. The observed Ba abundances at
$-4\lesssim{\rm [Fe/H]}< -3$ are particularly important as they
also are the key evidence requiring a much larger dilution mass 
in the regime where VMS are active in comparison with the regime 
of [Fe/H]~$\gtrsim -3$ where VMS are no longer active.

Unlike SNe II, VMS explode due to pair instability (e.g., 
Rakavy, Shaviv, \& Zinamon 1967; Bond, Arnett, \& Carr 1984; 
Glatzel, Fricke, \& El Eid 1985; Heger \& Woosley 2001). We do
not consider VMS as a source of $r$-process elements such as Ba.
On the other hand, small abundances of Ba have been observed at
$-4\lesssim{\rm [Fe/H]}< -3$ (e.g., McWilliam et al. 1995;
McWilliam 1998; Norris et al. 2001). This was previously
considered by us to be the result of contamination of the
VMS-enriched ISM by a small amount of SN II($H$) ejecta that
contains Ba but very little Fe (WQ). We now consider that 
some level of ``normal'' astration was present in the
regime where VMS were active and that the
abundances in the ISM at $-4\lesssim{\rm [Fe/H]}< -3$ consist of
contributions from VMS and SNe II($H$) and II($L$). 
As Ba is not considered to be
produced by VMS, we may attribute the Ba in a star with these
metallicities to an effective number $\tilde n_H$ of SNe II($H$)
and an effective number $\tilde n_L$ of SNe II($L$) that had
contributed to the ISM from which the star was formed:
\begin{equation}
\label{ba}
10^{\log\epsilon({\rm Ba})}=\tilde n_H\times
10^{\log\epsilon_H({\rm Ba})}+\tilde n_L\times
10^{\log\epsilon_L({\rm Ba})},
\end{equation}
where $\log\epsilon_H({\rm Ba})=-1.57$ or $\log\epsilon_L({\rm
Ba})=-0.47$ corresponds to the abundance of Ba in the ISM
resulting from a single SN II($H$) or II($L$) for a standard
dilution mass of $M_{\rm dil}^{\rm SNII}\approx 3\times
10^4\,M_\odot$ (QW). The lowest observed $\log\epsilon({\rm Ba})$
value is $-2.97$ for CS 22885-096 ([Fe/H]~$=-3.66$; Norris et al.
2001). Equation (\ref{ba}) gives $\tilde n_H=0.04$ if the Ba
abundance in this star was provided by a single SN II($H$) or
$\tilde n_L=3.16\times 10^{-3}$ for a single SN II($L$). We
consider that these low values of $\tilde n_H$ and $\tilde n_L$
require a much larger dilution mass for the SN II ejecta at
$-4\lesssim {\rm [Fe/H]}< -3$ compared with the standard dilution
mass at [Fe/H]~$\gtrsim -3$. This is in accord with the extremely
energetic explosions of VMS (e.g., Heger \& Woosley 2001) that
would occur along with SNe II at $-4\lesssim {\rm [Fe/H]}< -3$.
Specifically, the dilution mass $M_{\rm dil}^{\rm VMS}$ relevant
for these ultra-low metallicities is at least $M_{\rm dil}^{\rm
SNII}/0.04\approx 8\times 10^5\,M_\odot$ and may be as large as
$M_{\rm dil}^{\rm SNII}/ (3.16\times 10^{-3})\approx 9\times
10^6\,M_\odot$ (the difference in the dilution masses for
[Fe/H]~$<-3$ and [Fe/H]~$\gtrsim -3$ was considered earlier by
Qian \& Wasserburg (2001a) in discussing the $P$-inventory of O).

We now use the Ba abundances to estimate the SN II contributions
for stars with $-4\lesssim {\rm [Fe/H]}< -3$ in order to obtain
the ``pure'' VMS contributions. As explosions of VMS can disrupt
regions with a mass much larger than the standard dilution mass 
for SNe II, we assume that the overall mixing occurs over the 
region governed by VMS explosions. We also assume that 
$\tilde n_H/\tilde n_L=10$
for a star with $-4\lesssim {\rm [Fe/H]}< -3$ and use equation
(\ref{ba}) to estimate $\tilde n_L$ as
\begin{equation}
\label{nlba}
\tilde n_L=10^{\log\epsilon({\rm Ba})-\log\epsilon_{HL}({\rm Ba})}
=10^{\log\epsilon({\rm Ba})+0.22},
\end{equation}
where we have defined
\begin{equation}
\label{ehl}
10^{\log\epsilon_{HL}({\rm Ba})}\equiv
10\times 10^{\log\epsilon_H({\rm Ba})}+10^{\log\epsilon_L({\rm Ba})}
=10^{0.22}.
\end{equation}
The results presented below are not at all sensitive to the choice of
$\tilde n_H/\tilde n_L=10$ as 
the Ba yield of SNe II($L$) is $\approx 13$ times larger
than that of SNe II($H$) (see Table \ref{tabyphl}).
For an element E other than Ba, the fraction
$f_{HL}({\rm E})$ of this element in the star contributed by SNe
II($H$) and II($L$) is
\begin{equation}
\label{fhl}
f_{HL}({\rm E})\equiv\tilde n_L\times
10^{\log\epsilon_{HL}({\rm E})-\log\epsilon({\rm E})},
\end{equation}
where $\log\epsilon_{HL}({\rm E})$ is defined in a similar manner
to $\log\epsilon_{HL}({\rm Ba})$ and can be calculated from the
$\log\epsilon_H$ and $\log\epsilon_L^{\rm corr}$
values in Table \ref{tabyphl}. The remaining fraction
$[1 - f_{HL}({\rm E})]$ is then attributed to VMS:
\begin{equation}
\label{epsvl}
10^{\log\epsilon_{\rm VMS}({\rm E})}=[1-f_{HL}({\rm E})]\times
10^{\log\epsilon({\rm E})},
\end{equation}
where $\log\epsilon_{\rm VMS}({\rm E})$ represents the total VMS
contributions to the element E in the star. The above equation 
also applies to the $P$-inventory.

By using the $\log\epsilon_H$ and
$\log\epsilon_L^{\rm corr}$ values in Table
\ref{tabyphl} together with equations (\ref{nlba}), (\ref{fhl}), and
(\ref{epsvl}),  the $f_{HL}$ and $\log\epsilon_{\rm VMS}$ values
for Na to Ni and Sr are calculated from the data on four stars
with [Fe/H]~$=-3.98$ to $-3.61$ (Norris et al. 2001) and from the
$P$-inventory as determined at [Fe/H]~$=-3$. These results are
given in Table \ref{tabyvms}. The VMS yield patterns represented by
the $\log\epsilon_{\rm VMS}$ values are
also shown in Figure \ref{figvms}. For convenience of comparison,
the $\log\epsilon_{\rm VMS}$ values given in Table \ref{tabyvms} and
shown in Figure \ref{figvms} have been shifted from the original
calculated values to pass through the same Fe abundance of
$\log\epsilon_{\rm VMS}({\rm Fe})= \log\epsilon_L({\rm Fe})=5.03$.
It can be seen that the yield patterns of VMS derived from
the four stars and that derived from the $P$-inventory are almost 
identical to each other. The only exceptions are the anomalies at 
Na, Mg, and Si in CS 22949-037 as noted earlier and the varying 
production of Sr. Table \ref{tabyvms} also shows that VMS dominated
the production of Na to Ni at $-4\lesssim [{\rm Fe/H}]\lesssim -3$
and that SNe II($L$) contributed $\sim 3$--10\% of the
Fe at these metallicities.

The VMS yield patterns derived above may be compared with the 
SN II($L$) yield
pattern. Figure \ref{figvms} shows that relative to SNe II($L$),
VMS underproduce Na, Al, V, Cr, and Mn, overproduce Co, produce a
variable amount of Sr, but otherwise have an almost identical 
yield pattern. The Sr production by VMS is small or completely 
negligible based on the data for CD $-38^\circ 245$, CS 22172-002, 
and CS 22885-096, but may be somewhat larger than that by 
SNe II($L$) based on the data for CS 22949-037 and the
$P$-inventory. Thus there appears to be a strong hint for major
Sr production by some VMS.

\section{Model for General Chemical Evolution}
\label{secevol}

We have presented in \S\S\ref{secysn} and \ref{secyvms}
the results on SN
II($H$) and II($L$) yields and SN Ia and VMS yield patterns, all
of which appear to be rather well-defined. Based on these results,
we now discuss a model for general chemical evolution of the ISM
starting from a zero-metallicity state. We will extend the earlier
approach for [Fe/H]~$\gtrsim -3$ (e.g., Qian \& Wasserburg 2001a;
QW) to $[{\rm Fe/H}] < -3$. The production by VMS, and SNe II($H$),
II($L$), and Ia, as well as the difference in the dilution masses
associated with SNe II and VMS, will be taken into account. A full
approach must address the discreteness of the various production
events. We will first use a continuous approximation to illustrate
some essential features of the model and discuss the effects of
discrete events later. We use E to represent
an element other than Fe. We consider the following sequence
of evolution for the abundances of E and Fe in the ISM. Starting
from big bang debris with zero metallicity, i.e.,
$\log\epsilon({\rm E})=-\infty$ and [Fe/H]~$=-\infty$, the ISM was
first enriched in E and Fe by VMS and SNe II($H$) and II($L$).
When a metallicity of [Fe/H]~$\approx -3$ was reached, VMS ceased
occurring. At $-3\lesssim{\rm [Fe/H]}<-1$, only SNe II($H$) and
II($L$) provided chemical enrichment of the ISM, while at
[Fe/H]~$\gtrsim -1$, SNe II($H$), II($L$), and Ia all contributed. 
We are concerned here only with the contributions from VMS and
SNe II and Ia, but not with the contributions from low mass stars
such as asymptotic giant branch (AGB) stars. AGB stars would
make $s$-process contributions to the heavy elements above Fe at 
[Fe/H]~$\gtrsim -1$. Thus we restrict
our discussion to the elements from Na to Ni for [Fe/H]~$\leq 0$
and to the $r$-process elements (Sr and above) for [Fe/H]~$<-1$.

\subsection{Continuous Model without SN Ia Contributions}

We here consider a continuous model for evolution of elemental
abundances. This model is not applicable in regions where the
number of nucleosynthetic events is small and addition from
an individual event causes significant increment in the inventory
of an element. This approach does provide a good overall view
and in particular shows quantitatively that for example,
the evolution of Ba relative to Fe with a jump at [Fe/H]~$\approx -3$
is the result of the change from Fe production predominantly by VMS
to Fe production by SNe II($L$). The continuous model can be discussed
without knowing the dilution masses for [Fe/H]~$<-3$ and $\gtrsim -3$. 
However, as we have already 
demonstrated the change in the dilution mass for these two regimes
in \S\ref{secyvms}, we will incorporate this behavior in the
treatment here. The conclusion regarding the jump
in the evolution of Ba does not depend on the different dilution masses 
for [Fe/H]~$<-3$ and $\gtrsim -3$.

We first discuss the regime of [Fe/H]~$<-1$ that excludes SN Ia
contributions. Consider a homogeneous system of gas with a
time-dependent total number (H) of H atoms.
In the continuous approximation, the rate of change in the
abundance of E in this gas, $d({\rm E/H})/dt$, is (QW)
\begin{equation}
\label{eprod}
{d({\rm E/H})\over dt}={P_{\rm E}^{\rm VMS}(t)+
P_{\rm E}^H(t)+P_{\rm E}^L(t)\over({\rm H})},
\end{equation}
where $P_{\rm E}^{\rm VMS}(t)$, $P_{\rm E}^H(t)$, and $P_{\rm
E}^L(t)$ are the rates for production of E by VMS and SNe II($H$)
and II($L$), respectively. The production rates of individual
sources are proportional to the absolute yields and the
frequencies of these sources. Thus the terms such as $P_{\rm
E}^{\rm VMS}(t)/({\rm H})$ represent the products of the absolute
yields of each event and the frequency of occurrences per H atom
in the ISM for the individual sources. We consider that the
absolute yields of VMS and SNe II($H$) and II($L$) are fixed. We
further assume that the frequencies of these sources are
proportional to the total number (H) of H atoms in the gas but
with different proportionality constants for $0\leq t<t_*$ and
$t\geq t_*$, where $t_*$ is the time when [Fe/H]~$=-3$ was
reached. Then the production rates per H atom, denoted by
$\tilde{\cal{P}}_{\rm E}^{\rm VMS}$, $\tilde {\cal{P}}_{\rm E}^H$,
and $\tilde {\cal{P}}_{\rm E}^L$ for $0\leq t<t_*$ and by
${\cal{P}}_{\rm E}^H$ and ${\cal{P}}_{\rm E}^L$ for $t\geq t_*$,
are all constant. We use a similar notation for Fe. Note that
$\tilde {\cal{P}}_{\rm Fe}^H={\cal{P}}_{\rm Fe}^H=0$ as SNe
II($H$) produce no Fe (WQ). The production rates per H atom for
SNe II($H$) and II($L$) at $0\leq t<t_*$ (with the tilde symbol)
are related to those at $t\geq t_*$ (without the tilde symbol) by
the very different dilution masses with which the SN II ejecta is
mixed in these two epochs:
\begin{equation}
\label{rdil}
{\tilde {\cal{P}}_{\rm E}^H\over {\cal{P}}_{\rm E}^H}=
{\tilde {\cal{P}}_{\rm E}^L\over {\cal{P}}_{\rm E}^L}=
{\tilde {\cal{P}}_{\rm Fe}^L\over {\cal{P}}_{\rm Fe}^L}=
{M_{\rm dil}^{\rm SNII}\over M_{\rm dil}^{\rm VMS}}.
\end{equation}
It is found in \S\ref{secyvms} that the dilution mass
$M_{\rm dil}^{\rm VMS}$ for $0\leq t<t_*$ is larger than
the dilution mass $M_{\rm dil}^{\rm SNII}$ for $t\geq t_*$ 
by a factor of $\sim 25$--300.

With the notation introduced above, the rates of change in the
abundances of E and Fe in the gas are
\begin{equation}
\label{dedt} {d({\rm E/H})\over dt}=\left\{\begin{array}{ll}
\tilde{\cal{P}}_{\rm E}^{\rm VMS}+\tilde {\cal{P}}_{\rm E}^H+
\tilde {\cal{P}}_{\rm E}^L, & 0\leq t<t_*,\\ 
{\cal{P}}_{\rm E}^H+{\cal{P}}_{\rm E}^L, & t\geq t_*,
\end{array}\right.
\end{equation}
and
\begin{equation}
\label{dfedt}
{d({\rm Fe/H})\over dt}=\left\{\begin{array}{ll}
\tilde{\cal{P}}_{\rm Fe}^{\rm VMS}+\tilde {\cal{P}}_{\rm Fe}^L,
& 0\leq t<t_*,\\
{\cal{P}}_{\rm Fe}^L, &t\geq t_*.\end{array}\right.
\end{equation}
Eliminating $t$ in equations (\ref{dedt})
and (\ref{dfedt}), we obtain
\begin{equation}
\label{dedfe}
{d({\rm E/H})\over d({\rm Fe/H})}=
\left\{\begin{array}{ll}
(\tilde{\cal{P}}_{\rm E}^{\rm VMS}+\tilde {\cal{P}}_{\rm E}^H+
\tilde {\cal{P}}_{\rm E}^L)/
(\tilde{\cal{P}}_{\rm Fe}^{\rm VMS}+\tilde {\cal{P}}_{\rm Fe}^L),
& [{\rm Fe/H}]<-3,\\
({\cal{P}}_{\rm E}^H+{\cal{P}}_{\rm E}^L)/{\cal{P}}_{\rm Fe}^L,
& -3\leq [{\rm Fe/H}]<-1.
\end{array}\right.
\end{equation}
Equation (\ref{dedfe}) with the initial conditions 
${\rm (E/H)}={\rm (Fe/H)}=0$ at $t=0$ can be solved to give
\begin{equation}
\label{ehfeh}
({\rm E/H})=\left\{\begin{array}{ll}
[(\tilde{\cal{P}}_{\rm E}^{\rm VMS}+\tilde {\cal{P}}_{\rm E}^H+
\tilde {\cal{P}}_{\rm E}^L)/(\tilde{\cal{P}}_{\rm Fe}^{\rm VMS}+
\tilde {\cal{P}}_{\rm Fe}^L)]({\rm Fe/H}),
& [{\rm Fe/H}]<-3,\\
({\rm E/H})_*+[({\cal{P}}_{\rm E}^H+{\cal{P}}_{\rm E}^L)/
{\cal{P}}_{\rm Fe}^L]
[({\rm Fe/H})-({\rm Fe/H})_*], & -3\leq [{\rm Fe/H}]<-1,
\end{array}\right.
\end{equation}
where (E/H)$_*$ and (Fe/H)$_*$ correspond to the abundances
of E and Fe at [Fe/H]~$=-3$.

The result for [Fe/H]~$<-3$ in equation (\ref{ehfeh}) can be
rewritten as
\begin{equation}
\label{epsefea}
\log\epsilon({\rm E})=[{\rm Fe/H}]+\log\epsilon_\odot({\rm Fe})+
\log[(\tilde{\cal{P}}_{\rm E}^{\rm VMS}+\tilde {\cal{P}}_{\rm E}^H+
\tilde {\cal{P}}_{\rm E}^L)/
(\tilde{\cal{P}}_{\rm Fe}^{\rm VMS}+\tilde {\cal{P}}_{\rm Fe}^L)].
\end{equation}
Thus, in the continuous approximation, the evolution of E relative
to Fe at [Fe/H]~$<-3$ follows a straight line of unit slope on a
plot of $\log\epsilon({\rm E})$ as a function of [Fe/H]. The
position of this line on the plot is determined by the ratio of
the relevant total production rates per H atom for E and Fe.  The
result for $-3\leq {\rm [Fe/H]}<-1$ in equation (\ref{ehfeh}) in
the limit of ${\rm (E/H)}\gg{\rm (E/H)}_*$ and ${\rm
(Fe/H)}\gg{\rm (Fe/H)}_*$ can be approximated as
\begin{eqnarray}
\log\epsilon({\rm E})&\approx &[{\rm Fe/H}]+
\log\epsilon_\odot({\rm Fe})+\log[({\cal{P}}_{\rm E}^H+
{\cal{P}}_{\rm E}^L)/{\cal{P}}_{\rm Fe}^L]\nonumber\\
&\approx &[{\rm Fe/H}]+\log\epsilon_\odot({\rm Fe})
+\log[(\tilde {\cal{P}}_{\rm E}^H+\tilde {\cal{P}}_{\rm E}^L)/
\tilde {\cal{P}}_{\rm Fe}^L],
\label{epsefeb}
\end{eqnarray}
where equation (\ref{rdil}) is used in writing the last term.
This again represents a straight line of
unit slope. However, due to the  termination of VMS activities at
[Fe/H]~$=-3$ and the corresponding change in the ratio of the
total production rates per H atom for E and Fe, the line for
equation (\ref{epsefeb}) is offset from that for equation
(\ref{epsefea}) by
\begin{equation}
\label{jump}
\log\left[\left({\tilde {\cal{P}}_{\rm E}^H+
\tilde {\cal{P}}_{\rm E}^L\over\tilde {\cal{P}}_{\rm E}^{\rm VMS}+
\tilde {\cal{P}}_{\rm E}^H+\tilde {\cal{P}}_{\rm E}^L}\right)
\left({\tilde {\cal{P}}_{\rm Fe}^{\rm VMS}+\tilde {\cal{P}}_{\rm Fe}^L
\over\tilde {\cal{P}}_{\rm Fe}^L}\right)\right].
\end{equation}
Thus the full evolutionary trajectory
of $\log\epsilon({\rm E})$ follows a straight line of
unit slope up to [Fe/H]~$=-3$ and then asymptotically 
approaches another
straight line of unit slope. 
The sharpness of the transition from the initial
line to the asymptotic line can be assessed by
the slope of the evolutionary trajectory at [Fe/H]~$=-3$:
\begin{equation}
\label{slope}
\left.{d\log\epsilon({\rm E})\over d[{\rm Fe/H}]}\right|_*=
\left.{d({\rm E/H})\over d({\rm Fe/H})}\right|_*
{({\rm Fe/H})_*\over({\rm E/H})_*}=
\left({\tilde{\cal{P}}_{\rm E}^H+\tilde{\cal{P}}_{\rm E}^L\over
\tilde{\cal{P}}_{\rm E}^{\rm VMS}+\tilde {\cal{P}}_{\rm E}^H+
\tilde {\cal{P}}_{\rm E}^L}\right)
\left({\tilde{\cal{P}}_{\rm Fe}^{\rm VMS}+\tilde {\cal{P}}_{\rm Fe}^L
\over\tilde{\cal{P}}_{\rm Fe}^L}\right).
\end{equation}

Note that $\tilde{\cal{P}}_{\rm E}^{\rm VMS}=0$ for $r$-process
elements such as Ba. For these elements the shift in equation
(\ref{jump}) is $\log[(\tilde{\cal{P}}_{\rm Fe}^{\rm VMS}+\tilde
{\cal{P}}_{\rm Fe}^L)/ \tilde {\cal{P}}_{\rm Fe}^L]$ and the slope
in equation (\ref{slope}) is $(\tilde{\cal{P}}_{\rm Fe}^{\rm
VMS}+\tilde {\cal{P}}_{\rm Fe}^L)/ \tilde {\cal{P}}_{\rm Fe}^L$.
Table \ref{tabyvms} shows that SNe II($L$) contributed
$\sim 3$--10\% of the Fe in stars with [Fe/H]~$<-3$. So
$(\tilde{\cal{P}}_{\rm Fe}^{\rm VMS}+\tilde {\cal{P}}_{\rm Fe}^L)/
\tilde {\cal{P}}_{\rm Fe}^L\sim 10$--30. This results in a rather
dramatic jump of $\sim 1$--1.5~dex between the evolutionary tracks
of $\log\epsilon({\rm Ba})$ at [Fe/H]~$<-3$ and [Fe/H]~$>-2.5$. By
using the Ba and Fe yields of SNe II($H$) and II($L$) in Table
\ref{tabyphl}, the full trajectory of $\log\epsilon({\rm
Ba})$ is calculated from equation (\ref{ehfeh}) (see
\S\ref{subsecevol}) and shown in Figure \ref{figba} for
$(\tilde{\cal{P}}_{\rm Fe}^{\rm VMS}+\tilde
{\cal{P}}_{\rm Fe}^L)/ \tilde {\cal{P}}_{\rm Fe}^L=10$
(solid curve) and 20 (long-dashed curve). 
We recognize that the sparsity
of data at [Fe/H]~$< -3$ does not permit an accurate 
determination of the evolution in this region. We consider that 
this evolution may be represented by the corresponding
parts of the trajectories shown in Figure \ref{figba} as
the jump between the trends of the data at 
[Fe/H]~$<-3$ and [Fe/H]~$>-2.5$ is adequately reproduced.

\subsection{General Continuous Model}
\label{subsecevol}

The above continuous model can be extended to the general case
including SN Ia contributions to E and Fe in a straightforward
manner. For convenience of discussion, we designate the following
regions for evolution of E relative to Fe: (A) [Fe/H]~$<-3$ with
contributions to E and Fe from VMS and SNe II($H$) and II($L$),
(B) $-3\lesssim [{\rm Fe/H}] \lesssim -2.5$ and (C) $-2.5<
[{\rm Fe/H}]<-1$ with contributions from SNe II($H$) and II($L$),
and (D) [Fe/H]~$\gtrsim -1$ with contributions from SNe II($H$),
II($L$), and Ia. A  typical evolutionary trajectory of
$\log\epsilon({\rm E})$ in these four regions is shown in Figure
\ref{figevol}a (solid curve). As discussed earlier, the trajectory
in region A follows a straight line of unit slope determined by
the ratio of the total rates per H atom for production of E and Fe
by VMS and SNe II($H$) and II($L$). Due to the termination of VMS
activities at [Fe/H]~$\approx -3$ and the corresponding change in the
ratio of the total production rates per H atom for E and Fe, the
trajectory in region B shifts from the line in region A. Note that
the termination of VMS activities also greatly decreases the
dilution mass for the SN II ejecta to the standard value. So the
$\log\epsilon({\rm E})$ and [Fe/H] values produced by VMS and SNe
II($H$) and II($L$) up to [Fe/H]~$\approx -3$ are soon overwhelmed by the
additional contributions from SNe II($H$) and II($L$) in region B.
This typically occurs at [Fe/H]~$\sim -2.5$ when a few SNe II($L$)
have occurred. The trajectory in region C then follows the
straight line of unit slope determined by the ratio of the total
rates per H atom for production of E and Fe by SNe II($H$) and
II($L$). At [Fe/H]~$\sim -1$, SNe Ia began to occur with major Fe
additions. Thus the ratio of the total production rates per H atom
for E and Fe may change again at the onset of SN Ia contributions.
The trajectory in region D then may shift from the line in region
C and reach another asymptotic straight line of unit slope after a
transition zone.

In order to apply the general continuous model to the data on
various elements, we need to know the ratios of the total
production rates per H atom for E and Fe in regions A, B
(or C), and D:
\begin{mathletters}
\begin{eqnarray}
\gamma_{\rm A}({\rm E})&\equiv&(\tilde{\cal{P}}_{\rm E}^{\rm VMS}+
\tilde {\cal{P}}_{\rm E}^H+\tilde {\cal{P}}_{\rm E}^L)/
(\tilde{\cal{P}}_{\rm Fe}^{\rm VMS}+\tilde {\cal{P}}_{\rm
Fe}^L),\\ \gamma_{\rm BC}({\rm E})&\equiv&({\cal{P}}_{\rm E}^H+
{\cal{P}}_{\rm E}^L)/{\cal{P}}_{\rm Fe}^L= (\tilde {\cal{P}}_{\rm
E}^H+\tilde {\cal{P}}_{\rm E}^L)/ \tilde {\cal{P}}_{\rm Fe}^L,\\
\gamma_{\rm D}({\rm E})&\equiv&({\cal{P}}_{\rm E}^H+
{\cal{P}}_{\rm E}^L+{\cal{P}}_{\rm E}^{\rm I})/ ({\cal{P}}_{\rm
Fe}^L+{\cal{P}}_{\rm Fe}^{\rm I}),
\end{eqnarray}
\end{mathletters}
where ${\cal{P}}_{\rm E}^{\rm I}$ and ${\cal{P}}_{\rm Fe}^{\rm I}$
are the rates per H atom for production of E and Fe, respectively,
by SNe Ia. For regions B and C we may estimate 
$\gamma_{\rm BC}({\rm E})$ directly from the SN II($H$) and II($L$) 
yields in Table \ref{tabyphl} as
\begin{equation}
\gamma_{\rm BC}({\rm E})=[10\times 10^{\log\epsilon_H({\rm E})}+
10^{\log\epsilon_L({\rm E})}]/10^{\log\epsilon_L({\rm Fe})},
\end{equation}
where we have used the frequency ratio of 10:1 for SNe II($H$) and
II($L$) (e.g., QW). 
For region A the ratio $\gamma_{\rm A}({\rm E})$ can be
estimated from
\begin{equation}
{\gamma_{\rm A}({\rm E})\over\gamma_{\rm BC}({\rm E})}=
\left({\tilde{\cal{P}}_{\rm E}^{\rm VMS}+\tilde {\cal{P}}_{\rm
E}^H+ \tilde {\cal{P}}_{\rm E}^L\over\tilde{\cal{P}}_{\rm E}^H+
\tilde{\cal{P}}_{\rm E}^L}\right) \left({\tilde{\cal{P}}_{\rm
Fe}^L\over\tilde{\cal{P}}_{\rm Fe}^{\rm VMS} +\tilde
{\cal{P}}_{\rm Fe}^L}\right)= {f_{HL}({\rm Fe})\over f_{HL}({\rm
E})},
\end{equation}
where $f_{HL}({\rm E})$ and $f_{HL}({\rm Fe})$ are the fractions
of E and Fe at [Fe/H]~$<-3$ contributed by SNe II($H$) and II($L$)
with typical values given in Table \ref{tabyvms}. 

Now to estimate $\gamma_{\rm D}({\rm E})$ for region D, 
we must consider the rates per H atom for Fe production
by both SNe II($L$) and Ia, ${\cal{P}}_{\rm Fe}^L$ and
${\cal{P}}_{\rm Fe}^{\rm I}$. Once the ratio
${\cal{P}}_{\rm Fe}^{\rm I}/{\cal{P}}_{\rm Fe}^L$ is determined,
$\gamma_{\rm D}({\rm E})$ can be estimated from
\begin{equation}
\label{gamd}
\gamma_{\rm D}({\rm E})={\gamma_{\rm BC}({\rm E})+
({\cal{P}}_{\rm E}^{\rm I}/{\cal{P}}_{\rm Fe}^{\rm I})
({\cal{P}}_{\rm Fe}^{\rm I}/{\cal{P}}_{\rm Fe}^L)\over
1+({\cal{P}}_{\rm Fe}^{\rm I}/{\cal{P}}_{\rm Fe}^L)}
\end{equation}
by using the SN Ia yield pattern in Table \ref{tabyphl}.
We recognize that the progenitors of SNe II($L$) and Ia evolve very
differently and the ratio 
${\cal{P}}_{\rm Fe}^{\rm I}/{\cal{P}}_{\rm Fe}^L$ may not be constant
over Galactic history. For simplicity, we assume a constant
${\cal{P}}_{\rm Fe}^{\rm I}/{\cal{P}}_{\rm Fe}^L$ and estimate its 
value by making use of the timescales
associated with SN II($L$) and Ia activities. We have assumed
that SNe II($L$) are responsible for $\approx 1/3$ of the solar
Fe inventory and SNe Ia for the remaining fraction of $\approx 2/3$. 
Therefore, we have $P_{\rm Fe}^{\rm I}T_{\rm I}/
(P_{\rm Fe}^LT_{\rm II})\approx 2$, where $T_{\rm I}$
and $T_{\rm II}$ are the periods of SN Ia and II($L$)
production prior to SSF.
The ratio $T_{\rm I}/T_{\rm II}$ is less than unity and depends on the 
time required for a sufficient number of low mass stars to evolve in
binaries to provide SN Ia precursors. This time may be
$\sim 1$--$3\times 10^9$~yr, to be compared with
$T_{\rm II}\sim 10^{10}$~yr. We consider that $0.7 \lesssim 
T_{\rm I}/T_{\rm II}\lesssim 1$ and use
$P_{\rm Fe}^{\rm I}/P_{\rm Fe}^L=2.5$ in equation (\ref{gamd})
to estimate $\gamma_{\rm D}({\rm E})$.

Examples of the full evolutionary trajectories for the general
continuous model over the range of $-4 \leq [{\rm Fe/H}] \leq 0$ are
shown in Figure \ref{figmnco} for Mn and Co along with
the available data on these two elements. For
comparison we also show the same results in the conventional
[E/Fe] representation. The cusps in the curves at [Fe/H]~$= -3$
and $-1$ are due to the transitions where production in
one regime changes to that in the next as described above. The
intrinsic yield patterns of the different sources are unchanged. 
Note that the yield pattern of Mn, Fe, and Co for VMS is distinct from
that for SNe II($L$). As VMS turn off and SNe II continue with
increased production rates per H atom at [Fe/H]~$= -3$, the intrinsic
differences in the yield patterns of these two sources result in 
different changes in the ratios of the total production rates per H
atom and hence, in opposite trends for the evolution of Mn and Co 
in the region of $-4<[{\rm Fe/H}]<-2.5$.
These opposite trends as seen in the observational data have drawn
the attention of Nakamura et al. (1999). 
They sought to explain the observations by changing
the yields of the Fe group elements through different mass cuts
in ab initio models of SN II nucleosynthesis. 
In our approach, these trends are simply
explained by the existence of VMS at $-4\lesssim [{\rm Fe/H}]<-3$
with a yield pattern of Mn, Fe, and Co distinct from that of SNe
II($L$). The case of Cr is similar to Mn and can be explained in
the same manner by our approach.

The general continuous model can be used to discuss all the 
other elements in the group from Na to Ni as in the above 
examples for Mn and Co. Summary of the relevant data can be
found in the figures of McWilliam et al. (1995) and
Norris et al. (2001). A common feature for all these elements
is that the abundance ratios of other elements to Fe become
rather well defined when [Fe/H]~$\sim -2.5$ is reached. 
The typical spread in the abundance ratio of a particular element
to Fe is less than a factor of $\approx 3$ at a given metallicity
in the domain of [Fe/H]~$\gtrsim -2.5$.

\subsection{Effects of Discrete Events}

If there is more than one source for E and Fe as in the case of
Ba, the continuous approximation is good when a sufficient number
of production events of each kind have occurred in the ISM. If this
requirement is not satisfied, a discrete model should be used. The
Ba data in Figure \ref{figba} show that $\log\epsilon({\rm Ba})$
increases from $\approx -2.8$ at [Fe/H]~$\sim -4$ to $\approx
-1.8$ at [Fe/H]~$\sim -3$. So the number of Ba producing SNe
II($H$) and II($L$) increased by a factor of $\approx 10$ over the
range of $-4\lesssim{\rm [Fe/H]}<-3$ and the continuous
approximation should become  good as [Fe/H] approaches $-3$. At
[Fe/H]~$\gtrsim -3$, the dilution mass for the SN II ejecta decreased
greatly to the standard value and a single SN II($L$) would result
in [Fe/H]~$\approx -2.5$. So we expect that the continuous
approximation should also hold at [Fe/H]~$>-2.5$ after a few SNe
II($L$) had occurred with the standard dilution mass. The
evolutionary trends exhibited by the Ba data at [Fe/H]~$<-3$ and
[Fe/H]~$>-2.5$ are adequately described by two straight lines of
unit slope offset by $\sim 1$--1.5~dex as calculated from the
continuous model (see Figure \ref{figba}).

The effects of discrete events are most pronounced in region B of
$-3\lesssim [{\rm Fe/H}]\lesssim -2.5$ 
for $r$-process elements such as
Ba as illustrated schematically in Figure \ref{figevol}b. Due to
the enormous dilution mass for the SN II ejecta in region A, the
typical $\log\epsilon({\rm E})$ values for $r$-process elements in
this region are significantly below the ``base level'' resulting
from a single SN II($H$) with the standard dilution mass. This is
indicated by the downward arrows from the base level in Figure
\ref{figevol}b. After the cessation of VMS activities at
[Fe/H]~$\approx -3$, the standard dilution mass applies. A single SN
II($H$) would then correspond to point a and 10 SNe II($H$) to
point b in region B. As SNe II($H$) produce no Fe, the occurrence
of these events in region B is represented by the upward arrow
from point a to b. A single SN II($L$) with the standard dilution
mass would correspond to point c at the boundary between regions B
and C because it produces both Fe and e.g., Ba. As SNe II($H$) are 10
times more frequent than SNe II($L$), point d corresponding to 10
SNe II($H$) and a single SN II($L$) would lie on the trend line in
region C as calculated from the continuous model. For an SN
II($L$) frequency of $(10^8\ {\rm yr})^{-1}$ in a standard
dilution mass of ISM, region B corresponds to a period of
$\lesssim 10^8$~yr during which SNe II($H$) but not SNe II($L$) would 
occur with a high probability (WQ). 
Thus the continuous approximation does not apply to
region B and the trajectory in this region calculated from the
continuous model does not provide a physical description of the
data on e.g., Ba. The large scatter in e.g., $\log\epsilon({\rm
Ba})$ over the narrow range of $-3\lesssim{\rm [Fe/H]}\lesssim
-2.5$ is accounted for by the occurrence of $\sim 1$--10 SNe
II($H$) that produce Ba but no Fe in the discrete model (WQ).

\section{Discussion and Conclusions}
\label{secdis}

Based on a phenomenological model and the available data on
abundances in metal-poor stars, we have determined the yields of
the elements from Na to Ni for SNe II($H$) and II($L$) and the
yield patterns of the same elements for SNe Ia and VMS.
In a previous companion paper
(QW), we determined the yields of SNe II($H$) and 
II($L$) and the composition of the $P$-inventory for
$r$-process elements (Sr and above). The $P$-inventory of
$r$-process elements is now considered to be the result of
concomitant production by SNe II($H$) and II($L$) when VMS
were active. The $P$-inventory of non-neutron-capture
elements (possibly including Sr) is considered to be the
result of major production by VMS with small contributions
from SNe II($H$) and II($L$). The results presented here
and in QW show that the abundances of a large
number of elements (of both rapid-neutron-capture and 
non-neutron-capture origins) in any metal-poor star with
$-3\lesssim[{\rm Fe/H}]<-1$ can be almost quantitatively
determined from the observed [Fe/H] and 
$\log\epsilon({\rm Eu})$ values for the star by using the
three-component model including the $P$-inventory and the
contributions from SNe II($H$) and II($L$). 
A similar approach also applies
to stars with $-4\lesssim[{\rm Fe/H}]<-3$, for which
the contributions from VMS are explicitly included and
the contributions from SNe II($H$) and II($L$) are
determined by using the observed $\log\epsilon({\rm Ba})$ value 
as the index of $r$-process production.

By using the three-component model, it is shown
that SNe II($H$) control the abundances of all the heavy
$r$-process elements above Ba but contribute almost
nothing to the inventory of the elements between C and Ge.
The solar inventory of all the elements from O to V and the
solar $r$-process inventory of all the elements from Sr to Ba
are almost quantitatively accounted for by SNe II($L$).
The majority of the solar inventory of the Fe group elements
(Cr, Mn, Fe, Co, and Ni) is contributed by SNe Ia.
The metal inventory of the universe at [Fe/H]~$<-3$ was
dominated by the contributions from VMS.
Compared with SNe II($L$), VMS appear to have an almost
identical yield pattern but with a distinct deficiency
of Na, Al, V, Cr, and Mn and a distinct overproduction of
Co. All of these discrepant elements except for Cr are of 
odd atomic number. It is found that the metallicity of
[Fe/H]~$\approx -3$ represents the state when the occurrence
of VMS was essentially terminated.
While some formation of normal stars including SN II
progenitors and low mass stars occurred prior to the
achievement of [Fe/H]~$\approx -3$, the onset of major
formation of normal stars appears to be at this metallicity.
The rapid rise in the Ba abundance at [Fe/H]~$\approx -3$
(see Figure \ref{figba}) is due to the cessation of
VMS activities and the onset of more rapid normal star
formation. The mass of ISM with which the nucleosynthetic
products of VMS and SNe II are mixed in the domain of
[Fe/H]~$<-3$ is much larger (by a factor of 
$\sim 25$--300) than that for SNe II in the
domain of [Fe/H]~$\gtrsim -3$ where VMS ceased playing
any significant role. 

The above conclusions have implications
in four distinct areas: (1) as templates with which the
results of ab initio nucleosynthetic models of SNe II and VMS 
may be compared, 
(2) as predictions for the abundances relative to Fe 
in stars with ultra-low metallicities 
([Fe/H]~$<-3$), (3) as input for discussing the 
cosmological aspects of
condensation and dispersion of baryonic matter that
led to the formation of galaxies and other aggregations
of baryonic matter in the universe, and (4) as basis 
for considering the general chemical evolution of the universe.
In the following we will present a short summary on 
each of these issues based on the above findings.

\subsection{Implications for Nucleosynthetic Models of SNe II and VMS}

Based on the three-component model, we have shown in 
\S\S\ref{secysn} and \ref{secyvms} that the yield patterns of 
SNe II($H$) and II($L$) and VMS as inferred from
the observational data
exhibit remarkable regularity for a large number of elements.
The implications of these regular yield patterns
for nucleosynthetic models of SNe II and VMS are discussed below.

\subsubsection{SNe II($H$)}

It is commonly thought that in SNe II,
the elements below Si are produced
by hydrostatic burning during pre-SN evolution and those
above (including Si) are mostly produced by explosive burning
associated with the shock propagation. In order not to produce
a significant amount of the elements from Na to Ni, an SN II($H$)
must either have a pre-SN structure lacking substantial hydrostatic
burning shells or for some reason, have all the material below
the He burning shell fall back onto the central remnant. Stars
with masses of $\sim 8$--$11\,M_\odot$ have very thin shells
at the end of their lives and explode due to the collapse of
an O-Mg-Ne core instead of an Fe core (e.g., Nomoto 1984;
Hillebrandt, Nomoto, \& Wolff 1984; Mayle \& Wilson 1988). 
These stars could be
the progenitors for SNe II($H$). Mayle \& Wilson (1988) calculated
that an SN from O-Mg-Ne core collapse would only eject
$\sim 0.042\,M_\odot$ of matter and produce $\sim 0.002\,M_\odot$ 
of Fe. By assuming that $10^2$ SNe II($L$) must enrich a standard 
dilution mass of $\approx 3\times 10^4\,M_\odot$ with
$\approx 1/3$ of the solar Fe mass fraction
of $\approx 10^{-3}$, the Fe yield of an SN II($L$) is estimated
to be $\approx 0.1\,M_\odot$. This is $\sim 50$ times larger
than the Fe yield of an SN from O-Mg-Ne core collapse. Given
the extremely low total amount of ejecta, it is
reasonable to expect that SNe from O-Mg-Ne core collapse 
also do not produce any significant amount of other
elements below the Fe group. On the other hand, we note that
neutrino emission from the neutron star produced in any
core collapse would drive a small amount of material from the
neutron star crust. This neutrino-driven wind has been suggested
as a site of the $r$-process (e.g., Woosley et al. 1994) although
it remains to be seen if adequate conditions for the $r$-process
can be obtained in the wind (e.g., Qian \& Woosley 1996; 
Hoffman, Woosley, \& Qian 1997). We consider that SNe from
O-Mg-Ne core collapse may correspond to SNe II($H$) or at least
a subset (see also Wheeler, Cowan, \& Hillebrandt 1998). 
Further studies of these objects are extremely important
as most of the research on SNe II tends to focus on progenitors
with masses of 12--$40\,M_\odot$ that lead to explosion
from Fe core collapse.

Another possibility to accommodate the nucleosynthetic 
requirements of SNe II($H$) is that SNe II from Fe core collapse
would suffer severe fall back of processed material as emphasized
by Woosley \& Weaver (1995). It has been argued that SN 1997D,
which ejected only $0.002\,M_\odot$ of Fe, was such a case
(Benetti et al. 2001). The problem with such a scenario for
SNe II($H$) is that the severe fall back would be an obstacle
for ejection of the $r$-process material from the inner most
regions of the SN. In any case, we consider that many issues 
remain to be explored concerning the fall back and possible
production and ejection of $r$-process material in SNe from
Fe core collapse.

Given the validity of our model, we require that SNe II($H$)
to be the dominant source of the heavy $r$-process elements
above Ba but contribute very little of the other elements, 
especially
those between C and Ge. We also require SNe II($H$) to be the
frequent kind of SNe II. Regardless of the theoretical issues
involved in addressing the existence of such objects, clear
predictions can be made for the observational consequences of
SNe II($H$). First of all, Fe is mostly produced as the
radioactive $^{56}$Ni, whose decay powers the light curve of
an SN II at late stages. The lack of Fe production by 
SNe II($H$) means that the light curves of such events have
a very faint tail as in the case of SN 1997D. Based on
the frequent occurrences of SNe II($H$) in our model, it is 
likely that future systematic long-term observations of
SN II light curves may find a large number of cases like
SN 1997D. Another observational signature of SNe II($H$)
relies on the occurrences of these events in binaries.
A low mass companion would experience surface contamination
when an SN II($H$) explodes in a binary. As SNe II($H$)
produce mainly the heavy $r$-process elements above Ba but
little else, the surface of the low mass companion should
be highly enriched in the heavy $r$-process elements but
appear more or less normal in e.g., Na to Ni, when compared
with other stars with similar [Fe/H]. This exactly fits
the observations of CS 22892-052 as discussed in 
\S\ref{secysn}. We have also argued that CS 31082-001 is
another example of surface contamination by the SN II($H$)
ejecta in a binary (Qian \& Wasserburg 2001b) based on
the highly enriched abundances of heavy $r$-process elements
but very low value of [Fe/H]~$=-2.9$ for this star
(Cayrel et al. 2001). There are no available data 
on the elements in the group of Na to Ni other than Fe
for CS 31082-001. As the value of [Fe/H]~$=-2.9$ is almost 
the same as that for the $P$-inventory, we predict that future
observations of this star should detect abundances similar 
to the $P$-inventory for all the other
elements in the group of Na to Ni. The abundances of heavy
$r$-process elements observed in CS 31082-001 are extremely 
high and correspond to $n_H\sim 100$. Thus the observations
of Na to Ni in this star may be used to set a severe limit
on the production of these elements by SNe II($H$). We urge
that such observations be carried out soon.

\subsubsection{SNe II($L$)}

The SN II($L$) yields presented here are inferred from the data
on a star with [Fe/H]~$=-2.04$ corresponding to several 
contributing SNe II($L$). These yields almost
exactly follow the solar
abundance pattern up through V. This suggests that 
the SN II($L$) yields may be 
interpreted as the contributions from a single type of event and
that the yield pattern is constant, i.e., independent of
metallicity over Galactic history. 
The extremely regular SN II($L$) yields presented here may be 
compared with the yields calculated from nucleosynthetic 
models of SNe II, for which diversity appears to be the rule. 
Figure \ref{figylmod} compares the SN II($L$) yields 
(filled circles connected by solid curves) with
the yield patterns calculated from models
(Woosley \& Weaver 1995) of SNe II with
$20\,M_\odot$ progenitors of solar (triangles) and
0.01 times solar (squares) metallicities. 
The model yields have been shifted to
pass through the SN II($L$) yield of Mg in panel a and
that of Fe in panel b. These two choices of the reference
point are made to reflect that the elements below and above Si 
have different nucleosynthetic origins in SNe II.
It can be seen that the model yields are dependent on metallicity 
and that both model yield patterns only resemble the SN II($L$) 
yield pattern in local regions.

To emphasize
the regularity of SN II($L$) yields inferred from observations,
we show in Figure \ref{figsimg} the data on $\log({\rm Si/Mg})$
over the wide
range of $-4\lesssim [{\rm Fe/H}]\lesssim -1.3$. It can be seen that 
most ($\approx 75\%$) of the data are within a factor of 2 of 
Si/Mg~$=1\approx {\rm (Si/Mg)}_\odot$ and that there are no evidence
for high Si/Mg ratios of $> 3$. In contrast to this regularity,
the values of Si/Mg in most models (e.g., Woosley \& Weaver 1995) 
vary over a wide range of $\sim 0.1$--10. This variability
is expected as Mg is produced by hydrostatic burning but Si by
a mixture of hydrostatic and explosive burning. We emphasize that
averaging over many SNe II does not provide a viable way to
reconcile the diverse Si/Mg ratios in the models with
the uniform Si/Mg values exhibited by the data. This is because
the same uniformity of Si/Mg applies to the data at 
[Fe/H]~$\sim -2.5$,
which corresponds to Fe contributions from few SNe II($L$).

The modelling of explosive burning is extremely sensitive to
many uncertainties in our understanding of SNe II as emphasized
by Woosley \& Weaver (1995) and Thielemann, Nomoto, \& Hashimoto
(1996). This sensitivity is commonly explored by adopting various
mass cuts for the SN II ejecta. However, introduction of mass
cuts deviates from the ab initio approach and must rely on
empirical guidance such as the Fe yields inferred from SN II light 
curves. Note that the typical inferred Fe yields are 
$\sim 0.1\,M_\odot$ (e.g., Thielemann et al. 1996), which are 
consistent with the Fe yield estimated for SNe II($L$) in the 
preceding discussion on SNe II($H$). It is conceivable that by
choosing a mass cut to give the empirical Fe yield, SN II models
would give more convergent yield patterns similar to the SN II($L$)
yield pattern derived here from the observed abundances in 
metal-poor stars. A deeper issue then arises concerning the
physics and astrophysics underlying the mass cut.

\subsubsection{VMS}

We have shown that the yield pattern of VMS is also quite
regular. In particular, this pattern is almost identical to the 
SN II($L$)
yield pattern except for the distinct deficiency at Na, Al, V, Cr,
and Mn, the overproduction of Co, and the variable production of Sr
(see Figure \ref{figvms}). 
We will not discuss Sr any further. All the other discrepant
elements are of odd atomic numbers except for Cr. The underproduction 
of the elements of odd atomic numbers by VMS
is expected from the requirement of neutron excess
to produce these elements. 
During the evolution of an SN II progenitor,
significant weak interaction such as $\beta^+$ decay and electron 
capture would take place during hydrostatic burning to increase
the neutron excess. However, VMS encounter pair-instability after
core He burning and there is no stage of stable post-He burning to
increase the neutron excess (Heger \& Woosley 2001).  Thus, VMS
are expected to greatly underproduce the elements of odd atomic 
numbers relative to SNe II. However, the underproduction of Cr
of even atomic number, the overproduction of Co of odd atomic 
number, and the similar production of Sc of odd atomic 
number by VMS relative to SNe II($L$)
as found here are not readily explained.

Figure \ref{figyvmod} compares the VMS yield patterns inferred
from the data on four stars with $-4\lesssim {\rm [Fe/H]}<-3$
and that derived
from the $P$-inventory with the yields calculated from
two limiting models (Heger \& Woosley 2001) of VMS with He core 
masses of $70\,M_\odot$ (dashed curve) and
$130\,M_\odot$ (solid curve). The model yields have been
shifted to pass through $\log\epsilon({\rm Mg})=5.58$,
the location of most $\log\epsilon_{\rm VMS}({\rm Mg})$ values, 
in panel a and $\log\epsilon({\rm Fe})=5.03$,
the location of all $\log\epsilon_{\rm VMS}({\rm Fe})$ values,
in panel b. These two choices of the reference point are made
again to reflect the relative robustness of the yields for Mg
and Fe. It can be seen that below Ti, the model yields are quite 
robust and are a fair representation of the elements of even
atomic numbers. By contrast, the production of the elements above 
Ti in the models increases with the He core mass. It is also
evident that the production of the elements of odd atomic numbers
and Cr cannot be simply accounted for by the models
in a quantitative manner. Comparison with models for intermediate
He core masses does not change the basic conclusions.

The explosion energy in the VMS models of Heger \& Woosley (2001)
lies in the range of $\sim 10^{52}$--$10^{53}$~erg and increases
with the VMS mass. These values are larger than the typical SN II
explosion energy of $\approx 10^{51}$~erg by factors of
$\sim 10$--100. From detailed numerical studies, Thornton et al.
(1998) found that the total mass swept up by an SN remnant scales
with the explosion energy to a power of 6/7. This suggests a dilution
mass for VMS that is $\sim 7$--50 times the standard dilution mass 
of $M_{\rm dil}^{\rm SNII}\approx 3\times 10^4\,M_\odot$ for SNe II 
and is consistent with the lower end of the enormous dilution masses of 
$M_{\rm dil}^{\rm VMS}\sim 10^6$--$10^7\,M_\odot$ derived here for
VMS. Assuming that a single VMS gives [Fe/H]~$\sim -4$
in a dilution mass of $\sim 10^6\,M_\odot$,
we estimate that the Fe yield of VMS is $\sim 0.1\,M_\odot$.
This also implies that $\sim 10$ VMS had occurred in this dilution
mass prior to the cessation of
VMS activities at [Fe/H]~$\approx -3$
[Qian \& Wasserburg (2001a) concluded that many VMS are required
to provide the $P$-inventory of O].

\subsubsection{Summary}

From the detailed nucleosynthetic models of SNe II and VMS
we would expect there to be a wide range in the relative 
production of the elements from Na to Ni. However, we have
shown in \S\ref{secevol} that the evolution of the other elements 
relative to Fe over a wide range in [Fe/H]
can be adequately described by assuming 
fixed yield patterns of SNe II($L$) and VMS (see Figure
\ref{figmnco}). This poses a serious dilemma. One obvious solution 
would be that the sampling of the ISM by each star represents 
contributions from many SNe II and VMS whose average is strongly 
convergent. We well recognize that
mixing can hide grossly diverse sources. However, it is not
likely that the regular behavior exhibited by the data over 
the wide range in [Fe/H] can be all attributed to mixing.
For example, in the domain of $-2.5\lesssim {\rm [Fe/H]}\lesssim
-2$, the abundances of Na to Ni should be dominated by the 
contributions from only several SNe II($L$) and mixing would not
be effective. But the data in
Figure \ref{figmnco} show that the relative abundances of Mn
and Co to Fe already are well defined in this domain. 
The data on other elements
show the same behavior and can be found in the figures of
McWilliam et al. (1995) and Norris et al. (2001).

While a statistical equilibrium among the $\alpha$-nuclei from
$^{28}$Si to e.g., $^{52}$Fe, can be possibly at work to set
a fixed yield pattern of the corresponding elements,
there is no explanation, to our knowledge, of the yield patterns 
outlined here for SNe II($L$) and VMS. The observed constancy
of Si/Mg~$\approx 1$ over a wide range in [Fe/H] is in sharp
contrast to what is found in models of SNe II and VMS. We are 
pressed to
inquire whether or not a new paradigm for SN II and VMS models 
should be explored that would treat the possibility of
much more restrictive evolutionary paths for SNe II and VMS.
From the arguments presented here, the major efforts to model 
the solar abundance pattern by many 
sources with widely-varying yield patterns have not led to
a satisfactory conclusion. We recognize that ab initio 
calculations
are both complex and difficult. It is possible that some
unspecified dynamical effects involving partially quenched 
evolution
through high temperature and high density stages may be of 
importance in
the production of Mg and Si. The phenomenological approach used
here can in no way describe the detailed nuclear astrophysical
evolution of SNe II or VMS. However, we consider that the 
rather robust yield patterns of SNe II($L$) and VMS are quite 
directly obtained from the observational
data and we cannot identify any obvious
errors that would change the results. The only substantial variation 
is found for the VMS yields of Na, Mg, and Si from the data on
one star (CS 22949-037).
This lack of variation in SN II($L$) and VMS yields is remarkable 
in view of
the different data sources used. We do not consider that
this is plausibly attributable to some grand averaging. Instead,
there may be possible mechanisms that would give sharply
convergent yield patterns nearly independent of e.g., the mass or
metallicity of SN II progenitors. Certainly the observational
data should serve as a guide to theoretical exploration of
such mechanisms.

\subsection{Implications for Abundances at Ultra-low Metallicities}

Based on the model presented here, the abundances at ultra-low 
metallicities ([Fe/H]~$<-3$) can be described in terms of the
yield pattern of VMS and the yields of SNe II($H$) and II($L$).
For the elements from Na to Ni, the dominant contributions are 
from VMS. Thus, to first approximation, the observed abundances
at [Fe/H]~$<-3$ should follow the VMS yield pattern. 
A more detailed
study can be carried out by using the Ba abundance as the index
for $r$-process production to subtract the concomitant 
contributions from SNe II($H$) and II($L$). Such a study may
reveal more clearly the features at the elements of odd atomic
numbers and Cr as these elements 
receive more contributions from SNe II($L$).
All the $r$-process elements above Sr should be attributed to 
SNe II($H$) and II($L$). The abundances of these elements at
[Fe/H]~$<-3$ may be complicated by the effects of discrete events
and issues of the dilution mass. For example, a low mass star
might form in a region where an SN II($H$) just occurred but no
nearby VMS explosion occurred yet. In this case, the star would 
sample the enrichments by an SN II($H$) with a standard dilution
mass and thus show significant enhancements in the heavy 
$r$-process elements while maintaining an ultra-low [Fe/H].
Note that a star sampling the enrichments by an SN II($L$)
with a standard dilution mass would move out of the domain of
ultra-low metallicities as it would have [Fe/H]~$\approx -2.5$
(see Figure \ref{figevol}b). As SNe II($H$) are frequent,
future observations at [Fe/H]~$<-3$ may find stars with
unusually high abundances of heavy $r$-process elements.
In any case, given sufficient sensitivity, future observations 
may detect the heavy $r$-process elements such as Eu that were 
produced by SNe II($H$) concurrent with VMS activities
at ultra-low metallicities. It is evident that there is an 
urgent need for more extensive and precise observations of 
abundances in stars with metallicities well below $-3$.

\subsection{Implications for Formation of Galaxies and Chemical
Evolution of the Universe}

In considering broader implications of the results presented here,
it should be noted that the conclusions are only based on
observational data and a phenomenological model. No direct
considerations are made of cosmological models or the dynamics
and mechanisms associated with SN and VMS evolution. Time is,
for the most part, only considered
in terms of sequence. Increments in the inventory of elements
associated with newer generations of stars are taken as the measure
of time sequence. The other guiding rules are the general aspects of
stellar evolution.

The following general evolution sequence is inferred: (1)
formation of VMS made of big bang debris, (2) concomitant
formation at some level of a normal stellar population, (3)
``metal'' enrichment with overwhelming contributions from VMS
until a metallicity of [Fe/H]~$\approx -3$ was reached, (4)
cessation of VMS activities and onset of major formation of normal
stellar populations as marked by a sharp increase of contributions
from the frequent SNe II($H$) at [Fe/H]~$\approx -3$, (5)
continued chemical enrichment with dominant contributions from SNe
II($H$) and II($L$) until [Fe/H]~$\sim -1$ was reached, (6) onset
of SN Ia contributions at [Fe/H]~$\sim -1$ when a sufficient
number of low mass stars had evolved in binaries to serve as
precursors for SNe Ia, and (7) continued chemical enrichment with
contributions from SNe II($H$), II($L$), Ia and other evolved low
mass stars. In the regime of [Fe/H]~$<-3$ where VMS play a
predominant role, we infer that in aggregations of matter where
astration has begun, the explosion of a VMS redistributes and
mixes matter over a mass scale of $\sim 10^6$--$10^7\,M_\odot$.
This involves fresh nucleosynthetic products of the VMS and any SN
II debris in the region. Thus, when VMS activities cease at
[Fe/H]~$\approx -3$, the dilution mass with which nucleosynthetic
products are mixed drops by a few orders of magnitude from $\sim
10^6$--$10^7\,M_\odot$ for [Fe/H]~$<-3$ to the  value of
$\approx 3\times 10^4\,M_\odot$ associated with SNe II for
[Fe/H]~$\gtrsim -3$.

In typical cosmological models of hierarchical structure formation,
the condensation of dark matter is considered to provide 
the gravitational potential
wells in which some baryonic matter collects. The typical binding
energies of these wells correspond to escape velocities of $\sim
400$~km~s$^{-1}$. The VMS explosions are more than sufficient to
disrupt the baryonic condensates in such wells. We consider that
in the regime of [Fe/H]~$< -3$, the increases in ``metallicity''
are due to the sequence of local condensation of baryonic matter,
formation of VMS and some normal stars, and dispersion by VMS
explosion. There are then later reaggregations of some baryonic
matter and further enrichments through repeating the above
sequence. The cessation of VMS activities is due to the increase
in metallicity that favors formation of stars with a more normal
distribution in mass. This permits the aggregation of some matter
without disruption. Subsequent to the cessation of VMS activities
at [Fe/H]~$\approx -3$, large scale dispersion of baryonic matter
becomes very limited and large aggregation of baryonic matter that
leads to normal galactic structures and galactic chemical
evolution begins to occur. A recent study by Bromm et al. (2001)
indicates that substantial formation of a normal stellar
population could only occur when the metallicity is substantially
above $ 5\times 10^{-4}$ times the solar value.

In considering the domain where VMS and SNe II coexist (region A
in Figure 5), it is necessary to discuss the timing of events.
Based on the difficulty of forming stars with lower masses from
big bang debris (e.g., Bromm et al. 2001), the first objects formed
would have to be VMS. Upon explosion these would disrupt the original
baryonic aggregates formed in the potential wells of dark matter.
Reaggregation and repetition of such events would increase the
metallicity to allow formation of both VMS and lower mass stars.
Certainly some lower mass stars began to form by [Fe/H]~$\sim -4$.
Consider a ``cloud'' which contains VMS and SN II progenitors.
The VMS would explode first and disrupt the cloud. The subsequent
SN II explosions would disperse ejecta into a hot and tenuous medium
that does not permit any immediate star formation. After sufficient
cooling, the ejecta from the VMS and SNe II and the general medium 
would reaggregate under the influence of the dark matter 
potential. During this process, the nucleosynthetic 
products of VMS and SNe II would be effectively mixed over the entire 
mass of the baryonic aggregate prior to astration. Thus, the effective 
dilution mass for SNe II (much larger than the standard dilution mass) 
is the same as that for VMS in the regime where VMS are active. When
[Fe/H]~$\approx -3$ is reached, astration into lower mass stars becomes
very efficient and formation of VMS is truncated. The total timescale
over which normal astration becomes dominant is not known. Based on
considerations of data on damped Ly$\alpha$ systems, Wasserburg \&
Qian (2000b) estimated that this timescale was typically several
$10^9$~yr after the big bang. The question of reaggregation is a 
complex one as it involves condensation from ionized matter. It is
known that most of the baryonic matter resides in the ionized
intergalactic medium (IGM). This matter has not yet formed and may never
form galaxies.

Observations of damped Ly$\alpha$ systems over a wide range in
redshift ($z\approx 1.5$--4.5) show that the lowest [Fe/H]
observed is $\approx -2.7$ (e.g., Prochaska \& Wolfe 2000;
Prochaska, Gawiser, \& Wolfe 2001). 
The damped Ly$\alpha$ systems are thought
to represent proto-galaxies. The lower bound on [Fe/H] in these
systems was interpreted to reflect the transition to normal
astration at [Fe/H]~$\approx -3$ over an extended timescale after the
big bang (Wasserburg \& Qian 2000b). As this is the same effective
bound for a wide range of $z$, it follows that subsequent to the
cessation of VMS activities at [Fe/H]~$\approx -3$, most of the
baryonic matter remained dispersed in the universe to serve as a
reservoir for formation of proto-galaxies. Based on the model 
presented here, this matter should
exhibit the chemical enrichments corresponding to the
$P$-inventory that resulted from the integrated production by VMS
and some SNe II prior to the achievement of [Fe/H]~$\approx -3$.
Assuming the yields of Heger \&
Woosley (2001), Oh et al. (2001) have shown that a fraction $\sim
10^{-5}$--$10^{-4}$ of all baryonic matter must be processed
through VMS in order to account for the $P$-inventory at
[Fe/H]~$\approx -3$. Using the VMS luminosities of Bromm,
Kudritzki, \& Loeb (2001), they have further argued that this
amount of processing would provide $\sim 10$ photons per baryon at
energies sufficient to ionize H and He. This suggests that VMS may
be sufficient to explain the Gunn-Peterson effect (Gunn \&
Peterson 1965), which requires that most of the baryonic matter be
ionized. How effective the ionization by VMS may be for
[Fe/H] substantially below $-3$ is not yet explored. 
The $P$-inventory given here is
thus considered to represent the abundances in dispersed ionized
baryonic matter and should be compared with what is observed in 
e.g., the IGM. If an acceleration mechanism were to 
exist in the dispersed ionized baryonic matter, a cosmic ray 
component might be produced with a composition reflecting the 
$P$-inventory. 

It has long been recognized that the ratio E/Fe for the 
$\alpha$-elements such as Mg, Si, and Ca in the early Galaxy is
higher than the solar value by a factor of $\approx 3$.
This was due to the additional Fe production by SNe Ia
at later times (e.g., Tinsley 1980). Only $\approx 1/3$ of the solar
Fe inventory was produced by SNe II. The ratio E/Fe for
the $\alpha$-elements in the early Galaxy was considered to reflect 
the production of these elements by SNe II. We have shown that
the production of these elements at [Fe/H]~$<-3$ was dominated
by VMS. However, the yield pattern of Mg, Si, Ca, Ti, and Fe
for VMS is almost identical to that for SNe II($L$). Thus,
the ratio E/Fe for the $\alpha$-elements stays approximately
constant prior to the onset of SN Ia contributions although
the dominant production sources have changed at [Fe/H]~$\approx -3$ 
due to the cessation of VMS activities. The special status of 
[Fe/H]~$\approx -3$ is demonstrated by the sharp increase in Ba 
abundance at this metallicity (the same sharp increase also occurs at 
e.g., [Si/H]~$\approx -2.5$ as [Si/Fe]~$\approx 0.5$ at low [Fe/H]; 
see Figure 2 of Oh et al. 2001). The cosmological epoch prior to 
the achievement of [Fe/H]~$\approx -3$ remains to be explored.

\acknowledgments
We would like to thank Stan Woosley for education on nucleosynthesis
in supernovae and very massive stars and Volker Bromm for his 
penetrating questions. We also would like to thank
an anonymous referee for helpful comments. Y.Z.Q. acknowledges the 
hospitality of T-16 at the Los Alamos National Laboratory during
the initial phase of this work.
This work was supported in part by DOE grants DE-FG02-87ER40328 and
DE-FG02-00ER41149 (Y.Z.Q.) and by NASA grant NAG5-10293 (G.J.W.),
Caltech Division Contribution 8770(1081).

\clearpage

\figcaption{Data (filled circles: Ryan, Norris, \& Beers 1996; 
Norris et al. 2001; filled diamonds: McWilliam et al. 1995; 
McWilliam 1998; open squares: Burris et al. 2000; lower and upper
open triangles: HD 122563 and HD 115444, Westin et al. 2000; 
open circle: CS 22892-052, Sneden et al. 2000) on
$\log\epsilon({\rm Ba})$ as a function of [Fe/H]. 
The Ba yield of a single SN II($H$) is shown as the dotted line
and that of a single SN II($L$) is 1.1~dex above this. 
The straight dot-dashed line
represents the evolutionary trend for production only by SNe II($H$) 
and II($L$) with a frequency ratio of 10:1. The
extension of this line below [Fe/H]~$\sim -2.5$ corresponds to greater
than standard dilution masses but no other Ba or Fe sources.
The solid and long-dashed curves are for the continuous model of
evolution with VMS activities and the associated
large dilution mass at [Fe/H]~$< -3$. It is considered that
VMS produce no Ba but dominated the Fe production until the
cessation of VMS activities at [Fe/H]~$\approx -3$.
The solid and long-dashed curves assume that SNe II($L$) contributed
10\% and 5\%, respectively, of the Fe at [Fe/H]~$< -3$. The 
continuous approximation fails and a discrete model must be used in
the transition region between the two short-dashed lines
(see discussion of Figure \ref{figevol}).\label{figba}}

\figcaption{(a) Data on three stars (open squares: HD 122563
with [Fe/H]~$=-2.74$, filled circles connected with solid curves:
HD 115444 with [Fe/H]~$=-2.99$, Westin et al. 2000; asterisks:
CS 22892-052 with [Fe/H]~$=-3.03$, McWilliam et al. 1995) that
are used to infer the $P$-inventory and SN II($H$) yields. 
Note the remarkable similarity in the abundances of all the 
elements above C for these stars with widely varying Eu
abundances corresponding to SN II($H$) contributions for 
$n_H\approx 1$, 7 and 36. (b) The $P$-inventory (filled circles 
connected with solid curves) calculated from the data in (a).
The data on five stars (squares: CD $-24^\circ 17504$ with
[Fe/H]~$=-3.37$ and no Sr data, diamonds: CD $-38^\circ 245$ 
with [Fe/H]~$=-3.98$, triangles: CS 22172-002, asterisks: 
CS 22885-096, pluses: CS 22949-037; McWilliam et al. 1995 for Na 
and Norris et al. 2001 for all the other elements) are
shown for comparison. These data have been shifted from the 
observed abundances to pass through the same Fe abundance of
$\log\epsilon({\rm Fe})=\log\epsilon_P({\rm Fe})=4.51$ and
are almost indistinguishable from the $P$-inventory. There
appear to be some anomalies at Na, Mg, and Si for one star
(pluses: CS 22949-037), some spread in Mn, and a wide scatter 
in Sr.\label{figp}}

\figcaption{The SN II($L$) yields (filled circles connected with
solid curves) calculated from the data on HD 178443 
(McWilliam et al. 1995) by subtracting the small contributions
from the $P$-inventory and SNe II($H$). The solar abundance
pattern translated to pass through $\log\epsilon({\rm Mg})=
\log\epsilon_L({\rm Mg})=5.58$ (dashed curve) is shown for
comparison. With the exception of Al and Sc, the translated 
solar abundance pattern up through V coincides with the SN II($L$) 
yields. The differences for the Fe group elements (Cr, Mn, Fe, Co,
and Ni) are attributed to the sum of SN Ia contributions 
corresponding to the $\log\epsilon_{\odot,{\rm Ia}}$ values
in Table \ref{tabyphl}.\label{figyl}}

\figcaption{The VMS yield patterns calculated from the data on
four stars with [Fe/H]~$<-3$ (diamonds: CD $-38^\circ 245$,
triangles: CS 22172-002, asterisks: CS 22885-096, pluses:
CS 22949-037; McWilliam et al. 1995 for Na and Norris et al. 2001
for all the other elements) and that calculated from the
$P$-inventory at [Fe/H]~$=-3$ (open circles) by using Ba as the
index of SN II production. These patterns have been translated
to pass through the same Fe abundance of
$\log\epsilon({\rm Fe})=\log\epsilon_L({\rm Fe})=5.03$ for
comparison with the SN II($L$) yield pattern (filled circles
connected by solid curves). Note the general similarity between the
VMS and SN II($L$) yield patterns. However, VMS show some high 
production of Na, Mg, and Si from the data on one star (pluses:
CS 22949-037), and in all cases a significantly lower production
of Na, Al, V, Cr, and Mn and a significantly higher production
of Co. Except for CS 22885-096 (asterisks with no Sr assignment), 
the Sr in all the other three stars and in the $P$-inventory
is not explained by SN II coproduction with Ba. This is
considered as a strong hint of Sr production by some VMS.
\label{figvms}}

\figcaption{(a) Schematic illustration of continuous evolution
of $\log\epsilon({\rm E})$ as a function of [Fe/H] in four 
different regions (A, B, C, and D). The production sources in
each region are indicated. The cusps at [Fe/H]~$\approx -3$ and
$\sim -1$ are due to the transitions where production in one
region changes to that in the next (see text). (b) Schematic 
illustration of the effects of discrete events on the evolution
of $r$-process elements such as Ba. Due to the enormous dilution
mass associated with VMS activities in region A, the typical
abundances in this region are well below the base level 
resulting from a single SN II($H$) with the standard dilution
mass. The VMS activities cease at [Fe/H]~$\approx -3$ and the
dilution mass drops sharply to the standard value. Point a in
region B corresponds to a single SN II($H$) and point b to 10 
SNe II($H$). As SNe II($H$) produce no Fe, the occurrence of
these events in region B follow the upward arrow from point a 
to b. Point c corresponds to a single SN II($L$) and
point d to 10 SNe II($H$) and a single SN II($L$), all with
the standard dilution mass. The
continuous approximation fails and a discrete model must be
used in region B (see text).
\label{figevol}}

\figcaption{(a) Evolution of Mn relative to Fe for the 
continuous model (solid curves) is compared with the data
(filled circles: Ryan et al. 1996; Norris et al. 2001; 
filled diamonds: McWilliam et al. 1995; open squares: 
Gratton 1989; open triangles: Westin et al. 2000). The
asterisk corresponds to the solar data. The parameters 
used are $f_{HL}({\rm Mn})=0.3$, $f_{HL}({\rm Fe})=0.1$, and
${\cal{P}}_{\rm Fe}^{\rm I}/{\cal{P}}_{\rm Fe}^L=2.5$. No
attempt is made to fit the model to the data.
(b) Same as (a) but for Co (with open squares indicating
data from Gratton \& Sneden 1988 and with only one different
parameter $f_{HL}({\rm Co})=0.01$ for the model).
\label{figmnco}}

\figcaption{The SN II($L$) yield pattern (filled circles connected by
solid curves) is compared with the yield patterns calculated from
models (Woosley \& Weaver 1995) of SNe II with $20\,M_\odot$ 
progenitors of solar (triangles) and 0.01 times solar
(squares) metallicities. The model yields have been
shifted to pass through $\log\epsilon({\rm Mg})=
\log\epsilon_L({\rm Mg})=5.58$ in panel a
and $\log\epsilon({\rm Fe})=\log\epsilon_L({\rm Fe})=5.03$ 
in panel b. The solar abundance pattern translated to pass through
$\log\epsilon({\rm Mg})=5.58$ (dashed curve) is also shown in
panel a.\label{figylmod}}

\figcaption{The observed $\log{\rm (Si/Mg)}$ values over the range 
of $-4\lesssim [{\rm Fe/H}]\lesssim -1.3$ (filled circles:
Ryan et al. 1996; Norris et al. 2001; filled diamonds: 
McWilliam et al. 1995; open squares: Gratton \& Sneden 1988; 
open triangles: Westin et al. 2000). 
The solid line represents the solar value of 
$\log{\rm (Si/Mg)}_\odot=-0.03$. Note that 
almost all ($\approx 75\%$) of the data lie within $\pm 0.3$ dex
(between the dashed lines) of $\log{\rm (Si/Mg)}=0$. 
There is no evidence of very high Si/Mg ratios.
\label{figsimg}}
 
\figcaption{The VMS yield patterns derived from the data on four
stars with
[Fe/H]~$<-3$ and that derived from the $P$-inventory at
[Fe/H]~$=-3$ (same symbols as in Figure \ref{figvms}) are
compared with the yield patterns calculated from models
(Heger \& Woosley 2001) of VMS with He core masses of
$M_{\rm He}=70\,M_\odot$ (dashed curve) and
$130\,M_\odot$ (solid curve). The model yields have been
shifted to pass through $\log\epsilon({\rm Mg})=5.58$ in panel a
and $\log\epsilon({\rm Fe})=5.03$ in panel b.\label{figyvmod}}

\clearpage
\begin{deluxetable}{ccrrrrrrrr}
\tabletypesize{\scriptsize}
\tablecaption{$P$-Inventory, SN II($H$) and II($L$) Yields, and
SN Ia Yield Pattern\label{tabyphl}}
\tablewidth{0pt}
\tablecolumns{10}
\tablehead{
\colhead{$Z$\tablenotemark{a}}&\colhead{E}&
\colhead{$\log\epsilon_P({\rm E})$}&
\colhead{$\log\epsilon_H({\rm E})$}&
\colhead{$\log\epsilon_L({\rm E})$}&
\colhead{$\log\epsilon_\odot({\rm E})$}&
\colhead{$\alpha_{\odot,{\rm II}}({\rm E})$}&
\colhead{$\log\epsilon_L^{\rm corr}({\rm E})$}&
\colhead{$\alpha_{\odot,{\rm II}}^{\rm corr}({\rm E})$}&
\colhead{$\log\epsilon_{\odot,{\rm Ia}}({\rm E})$}\\
\colhead{(1)}&\colhead{(2)}&\colhead{(3)}&\colhead{(4)}&\colhead{(5)}&
\colhead{(6)}&\colhead{(7)}&\colhead{(8)}&\colhead{(9)}&\colhead{(10)}
}
\startdata
\sidehead{Group 1:}
6&C&$<5.46$&?&?&8.56&?&?&?&$-\infty$\\
7&N&6.30&$\lesssim 4.87$&?&8.05&?&?&?&$-\infty$\\
8&O&6.60&$\lesssim 5.17$&6.93&8.93&1.00&6.93&1.00&$-\infty$\\
\sidehead{Group 2:}
11&Na&3.34&$\lesssim 1.91$&4.24&6.33&0.81&4.33&1.00&$-\infty$\\
12&Mg&5.13&$\lesssim 3.70$&5.62&7.58&1.09&5.58&1.00&$-\infty$\\
13&Al&3.12&$\lesssim 1.69$&3.90&6.47&0.27&3.90&0.27&$-\infty$\\
14&Si&5.02&$\lesssim 3.59$&5.47&7.55&0.84&5.55&1.00&$-\infty$\\
20&Ca&3.75&$\lesssim 2.32$&4.40&6.36&1.10&4.36&1.00&$-\infty$\\
21&Sc&0.28&$\lesssim -1.15$&0.77&3.10&0.46&0.77&0.46&$-\infty$\\
22&Ti&2.43&$\lesssim 1.00$&2.87&4.99&0.76&2.99&1.00&$-\infty$\\
23&V&1.15&$\lesssim -0.28$&1.87&4.00&0.74&2.00&1.00&$-\infty$\\
24&Cr&2.40&$\lesssim 0.97$&3.17&5.67&0.32&3.17&0.32&5.50\\
25&Mn&1.90&$\lesssim 0.47$&2.56&5.39&0.15&2.56&0.15&5.32\\
26&Fe&4.51&$\lesssim 3.08$&5.03&7.51&0.33&5.03&0.33&7.34\\
27&Co&2.24&$\lesssim 0.81$&2.05&4.92&0.13&2.05&0.13&4.86\\
28&Ni&3.25&$\lesssim 1.82$&3.50&6.25&0.18&3.50&0.18&6.16\\
\sidehead{Group 3:}
29&Cu&0.53&$\lesssim -0.90$&?&4.21&?&?&?&?\\
30&Zn&1.86&$\lesssim 0.43$&?&4.60&?&?&?&?\\
32&Ge&$<-0.05$&?&?&3.41&?&?&?&$-\infty$\\
\sidehead{Group 4:}
38&Sr&0.13&$-1.30$&0.35&2.90&0.34&0.35&0.34&$-\infty$\\
56&Ba&$\approx -1.80$&
$-1.57$&$-0.47$&2.13&0.45&$-0.47$&0.45&$-\infty$\\
\enddata
\tablecomments{Results are given for four groups of elements
with the focus on groups 2 and 4.
The $P$-inventory in col. (3) and SN II($H$) yields in
col. (4) are calculated from the data shown in Figure \ref{figp}a
for groups 1--3. The SN II($L$) yields in
col. (5) are calculated from col. (3) and the data on HD 178443 
(McWilliam et al. 1995) for group 2 and that for O is calculated by
attributing the solar O inventory to $10^2$ SNe II($L$). 
Col. (6) gives the solar inventory as measured in the photosphere
(Anders \& Grevesse 1989).
Col. (7) gives the fraction of
the solar inventory of an element E contributed by SNe II as
calculated from cols. (4)--(6).
The corrected values in cols. (8) and (9) only reflect small
adjustments for those elements with 
$\alpha_{\odot,{\rm II}}({\rm E})\approx 1$.
The SN Ia yield pattern is represented by the part of the solar 
inventory contributed by SNe Ia in column (10) as calculated from
cols. (6) and (9).
The results for group 4 are taken from QW but with 
$\log\epsilon_P({\rm Ba})$ estimated from the data at
[Fe/H]~$\lesssim -3$ (see Figure \ref{figba}).}
\tablenotetext{a}{Atomic number.}
\end{deluxetable}

\clearpage
\begin{deluxetable}{ccrrrrrrrrrrrrrr}
\tabletypesize{\scriptsize}
\rotate
\tablecaption{VMS Yield Patterns\label{tabyvms}}
\tablewidth{0pt}
\tablehead{\colhead{}&\colhead{}&\multicolumn{2}{c}{CD $-38^\circ245$}&
\colhead{}&\multicolumn{2}{c}{CS 22172-002}&\colhead{}&
\multicolumn{2}{c}{CS 22885-096}&\colhead{}&
\multicolumn{2}{c}{CS 22949-037}&\colhead{}&
\multicolumn{2}{c}{$P$-inventory}\\
\cline{3-4}\cline{6-7}\cline{9-10}\cline{12-13}\cline{15-16}
\colhead{$Z$\tablenotemark{a}}&\colhead{E}&
\colhead{$f_{HL}({\rm E})$}&
\colhead{$\log\epsilon_{\rm VMS}({\rm E})$}&\colhead{}&
\colhead{$f_{HL}({\rm E})$}&
\colhead{$\log\epsilon_{\rm VMS}({\rm E})$}&\colhead{}&
\colhead{$f_{HL}({\rm E})$}&
\colhead{$\log\epsilon_{\rm VMS}({\rm E})$}&\colhead{}&
\colhead{$f_{HL}({\rm E})$}&
\colhead{$\log\epsilon_{\rm VMS}({\rm E})$}&\colhead{}&
\colhead{$f_{HL}({\rm E})$}&
\colhead{$\log\epsilon_{\rm VMS}({\rm E})$}
}
\startdata
11&Na&0.42&3.53&&\nodata&\nodata&&0.14&3.56&&$<0.01$&5.87&&0.26&3.77\\
12&Mg&0.11&5.55&&0.07&5.46&&0.02&5.62&&0.02&6.36&&0.07&5.66\\
13&Al&0.49&2.97&&0.36&2.88&&0.13&3.16&&0.23&3.49&&0.16&3.60\\
14&Si&0.15&5.34&&0.06&5.44&&0.03&5.51&&0.03&6.15&&0.09&5.54\\
20&Ca&0.22&3.96&&0.11&3.99&&0.04&4.15&&0.11&4.33&&0.11&4.26\\
21&Sc&0.12&0.69&&0.07&0.64&&0.03&0.77&&0.11&0.75&&0.08&0.80\\
22&Ti&0.14&2.82&&0.08&2.80&&0.05&2.67&&0.14&2.84&&0.10&2.95\\
23&V&$>0.09$&$<2.04$&&\nodata&\nodata&&$>0.04$&$<1.82$&&
             $>0.07$&$<2.19$&&0.19&1.62\\
24&Cr&0.44&2.34&&0.21&2.48&&0.10&2.57&&0.36&2.49&&0.15&2.89\\
25&Mn&0.51&1.59&&0.29&1.69&&0.07&2.10&&0.40&1.82&&0.12&2.40\\
26&Fe&0.10&5.03&&0.05&5.03&&0.03&5.03&&0.11&5.03&&0.09&5.03\\
27&Co&0.01&2.97&&0.01&3.03&&$<0.01$&3.04&&0.01&3.06&&0.02&2.79\\
28&Ni&0.07&3.66&&0.03&3.80&&0.01&3.96&&0.05&3.86&&0.05&3.79\\
38&Sr&0.54&$-0.57$&&0.85&$-1.58$&&1.10&$-\infty$&&0.09&0.53&&0.05&0.67\\
\enddata
\tablecomments{Results are calculated from the data on the
four listed stars with [Fe/H]~$<-3$ (McWilliam et al. 1995 for Na;
Norris et al. 2001 for all the other elements) and from the
$P$-inventory as determined for [Fe/H]~$=-3$. The parameter
$f_{HL}({\rm E})$ is the fraction of an element E in a star or in
the $P$-inventory contributed by SNe II($H$) and II($L$).
The VMS yield pattern is represented by 
$\log\epsilon_{\rm VMS}({\rm E})$ as calculated from $f_{HL}({\rm E})$
and the observed $\log\epsilon({\rm E})$ in a star or
$\log\epsilon_P({\rm E})$. For convenience of comparison, the
$\log\epsilon_{\rm VMS}$ values have been shifted to pass through the
same Fe abundance of
$\log\epsilon_{\rm VMS}({\rm Fe})=\log\epsilon_L({\rm Fe})=5.03$.}
\tablenotetext{a}{Atomic number.}
\end{deluxetable}
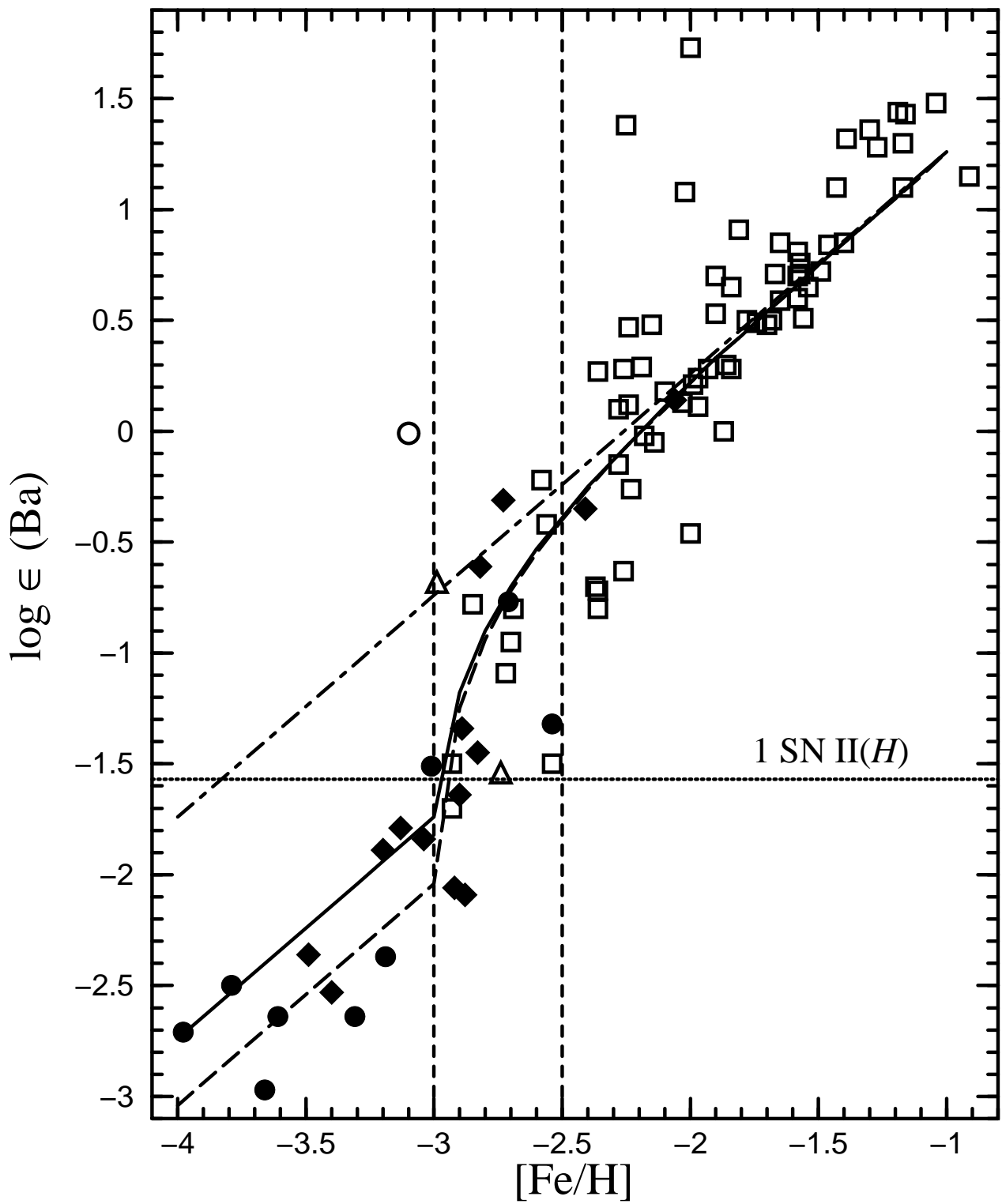
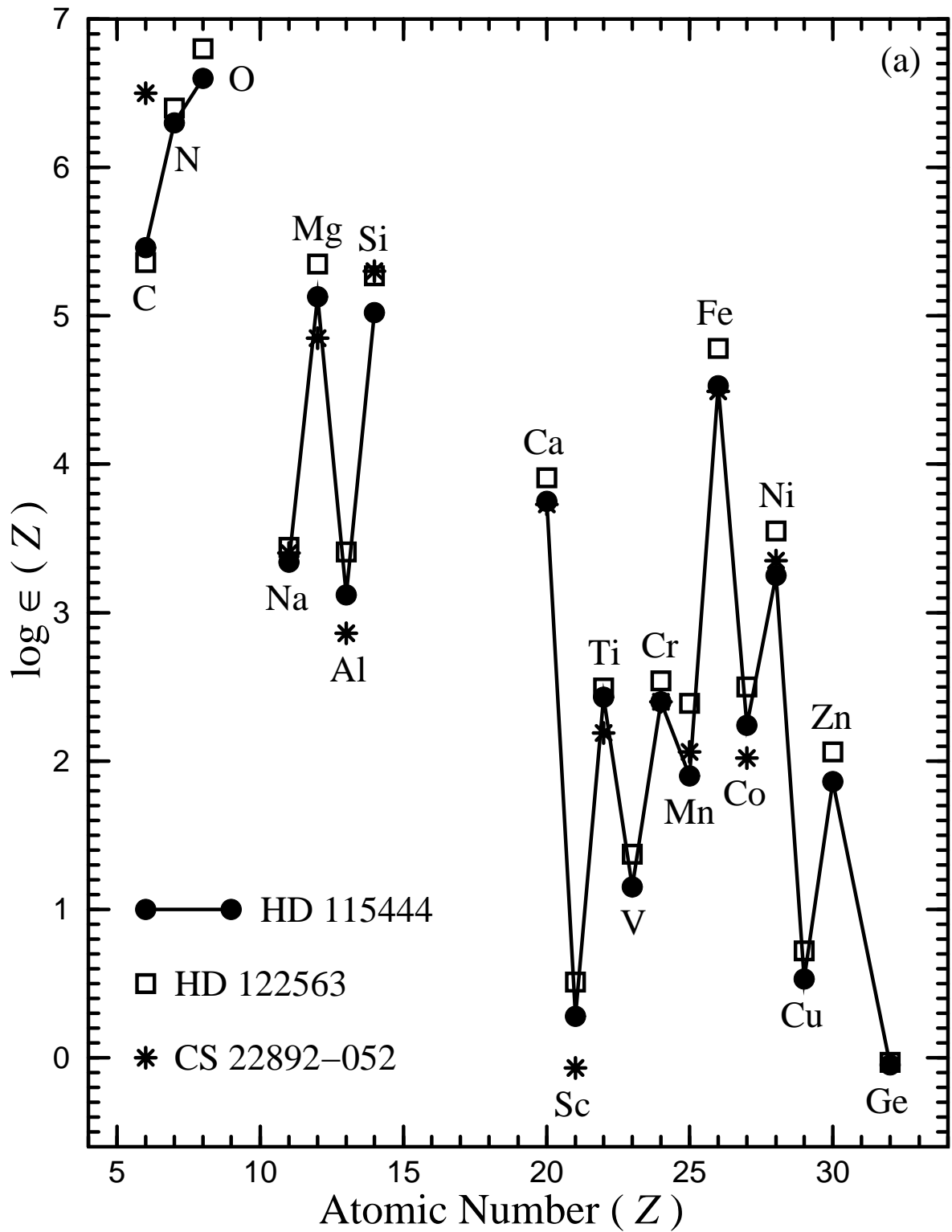
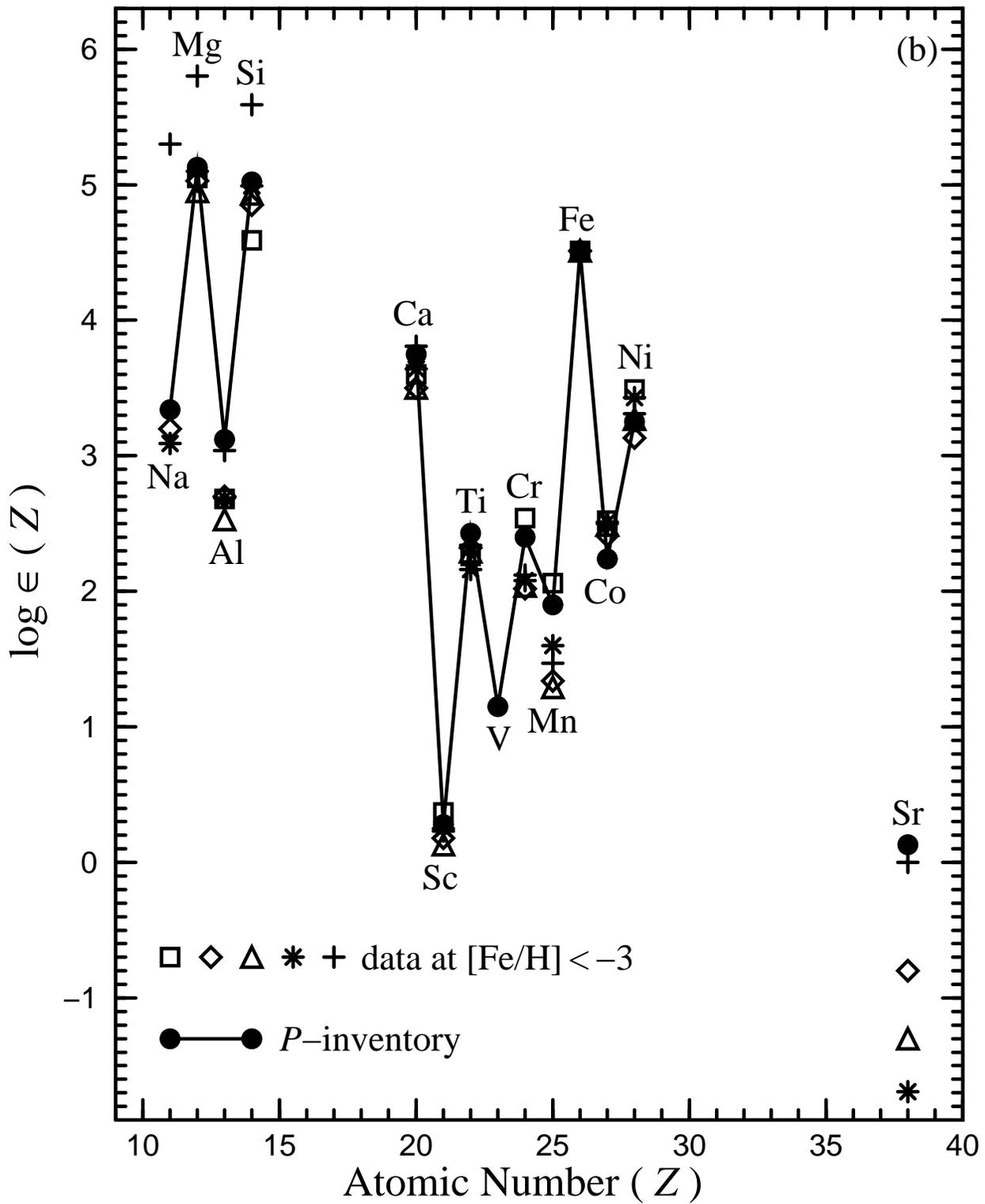
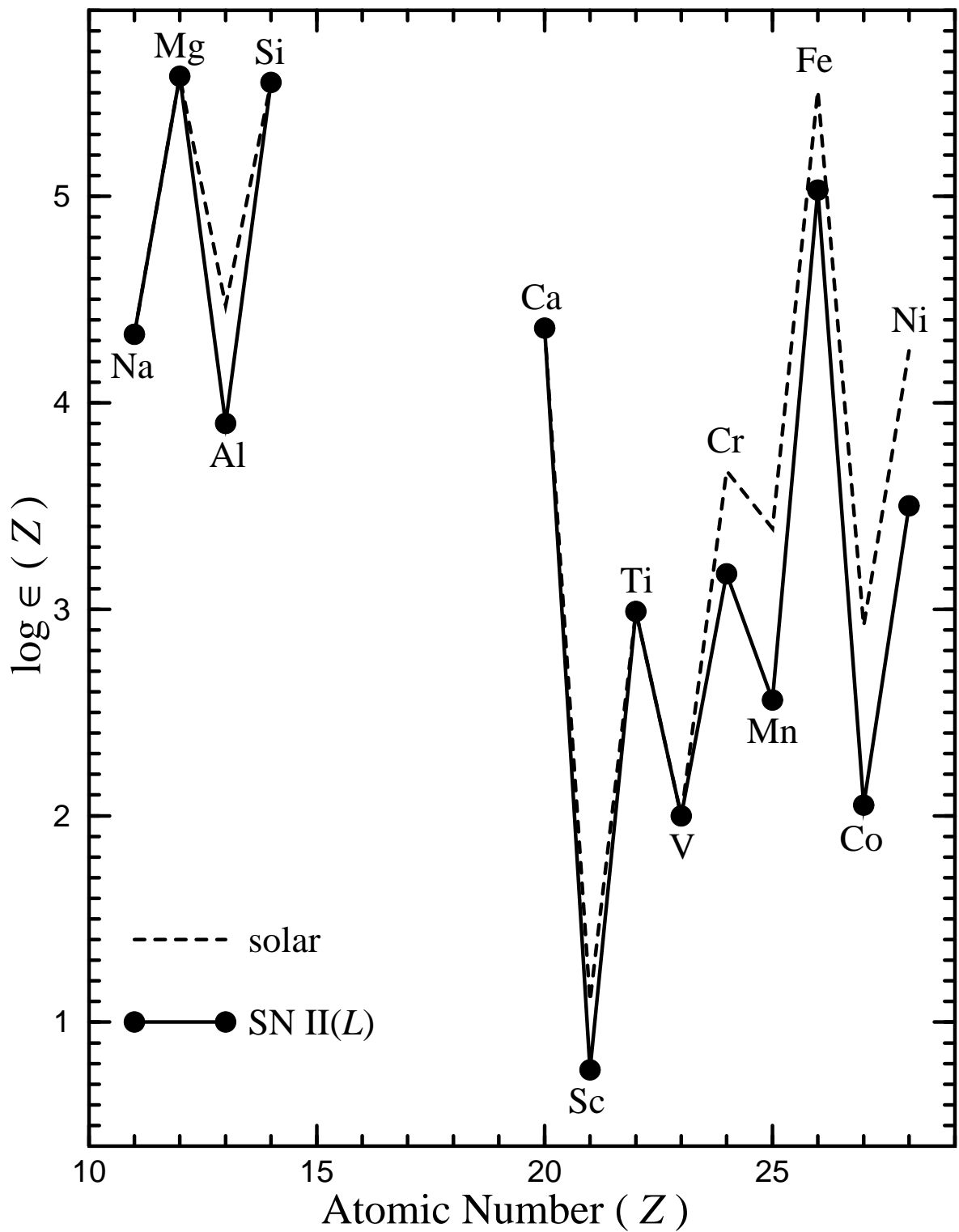
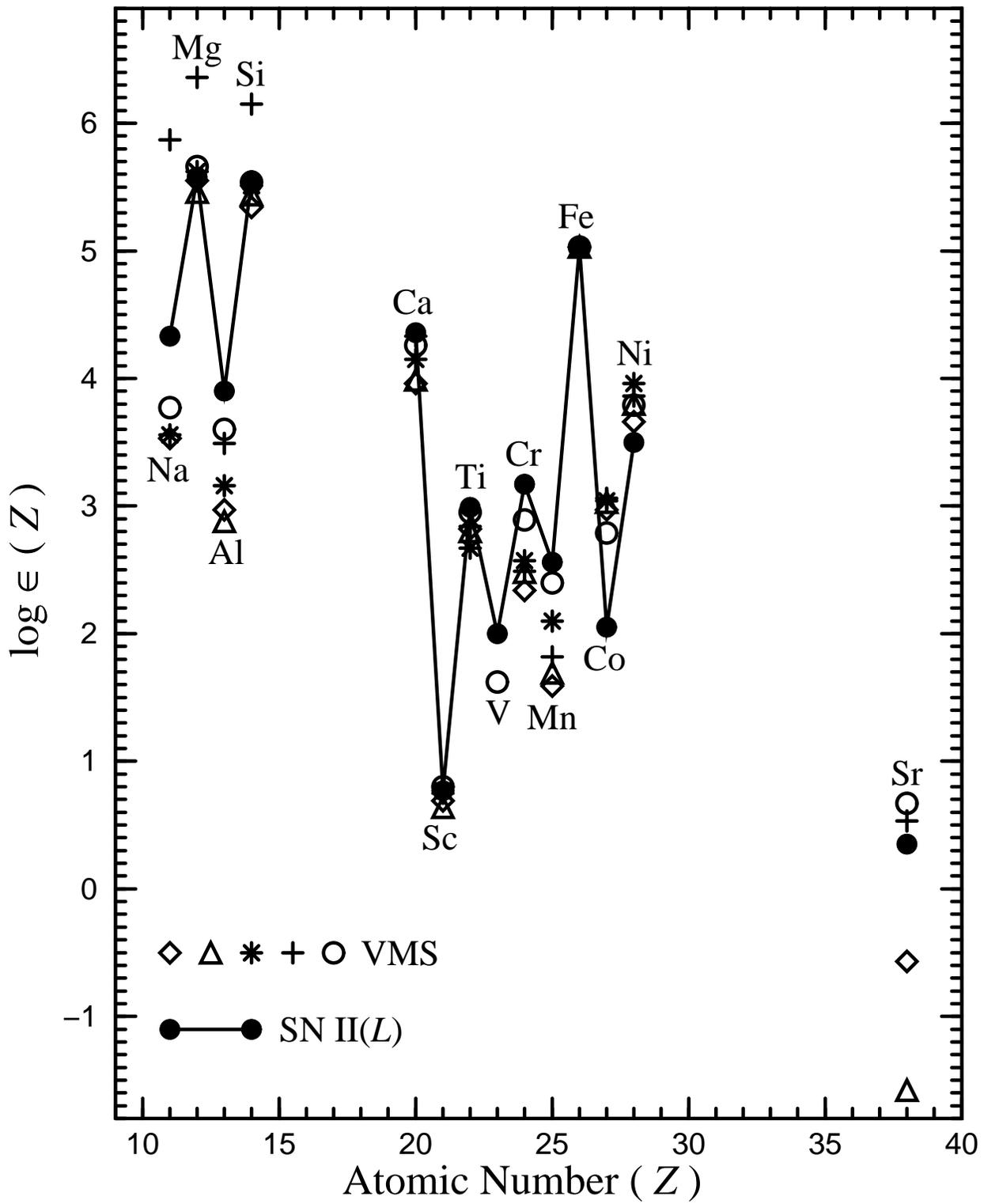
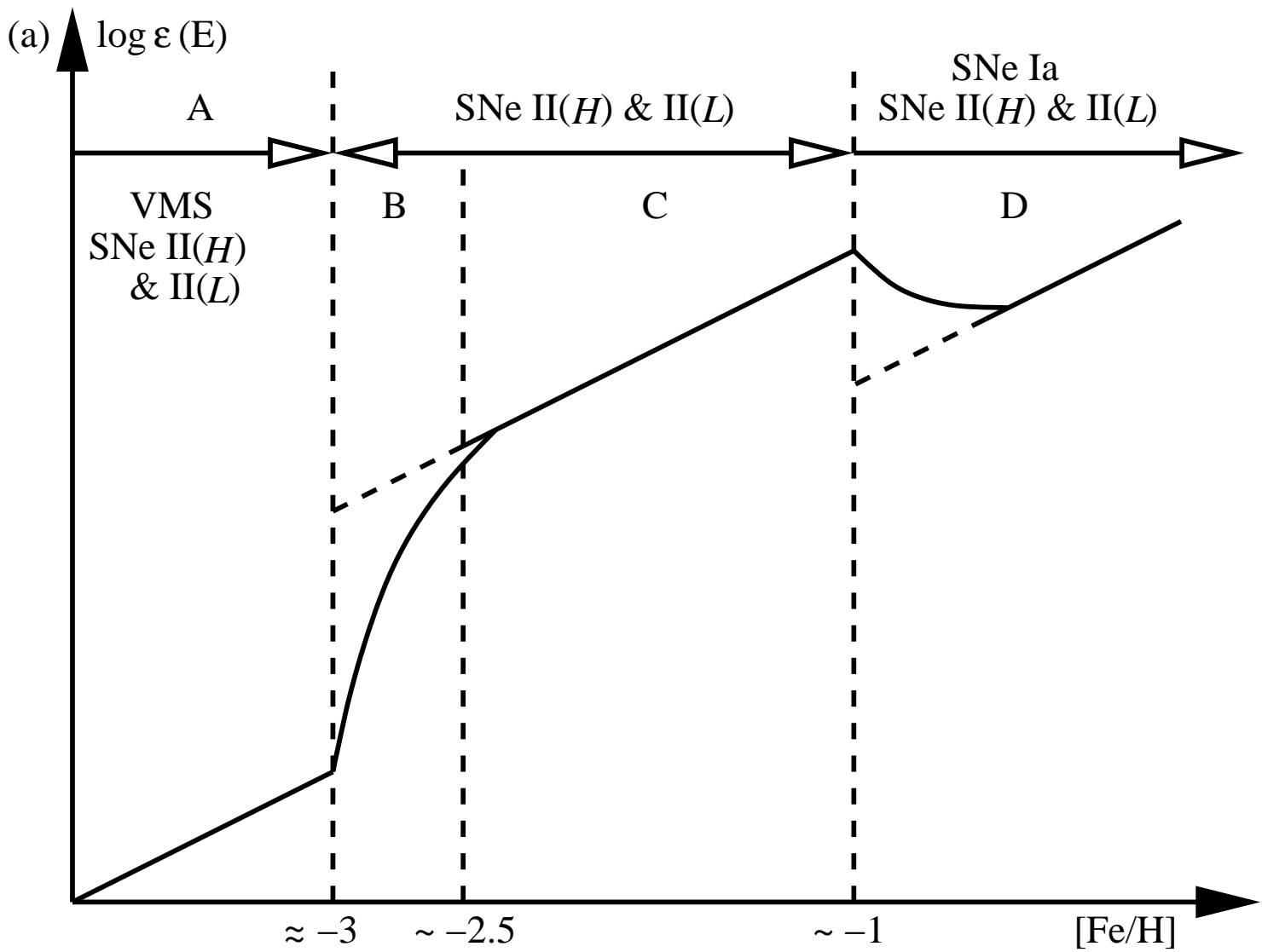
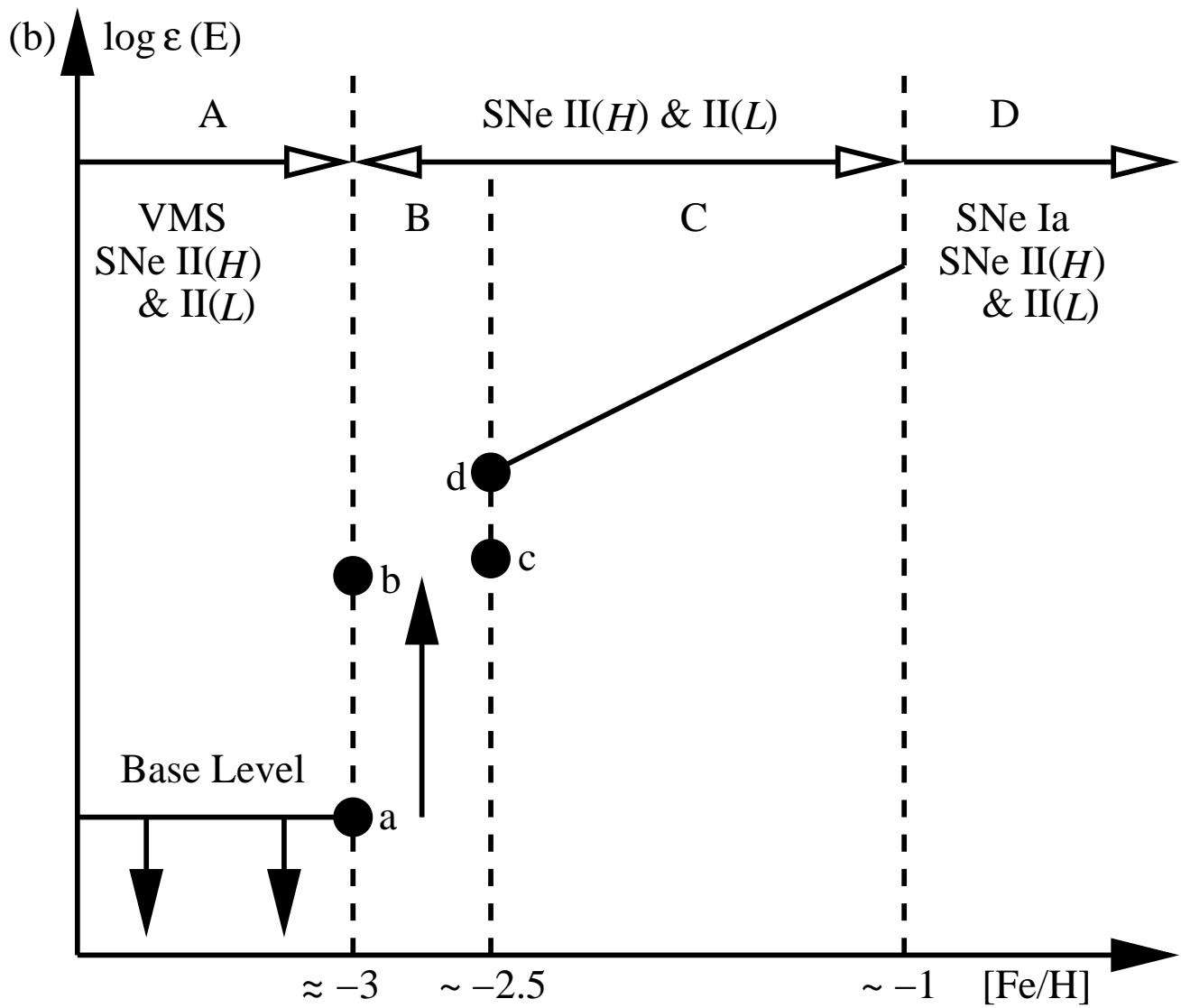
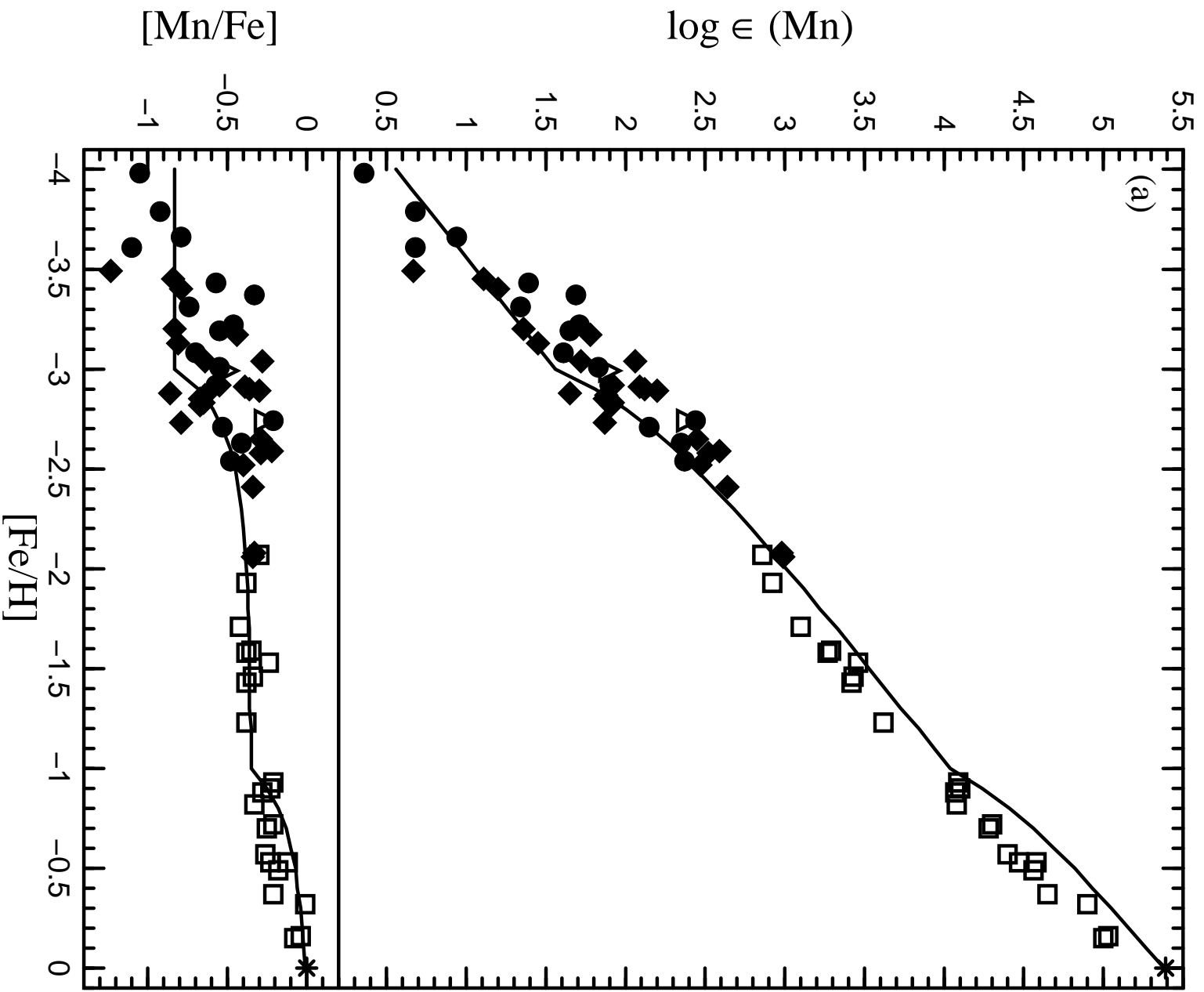
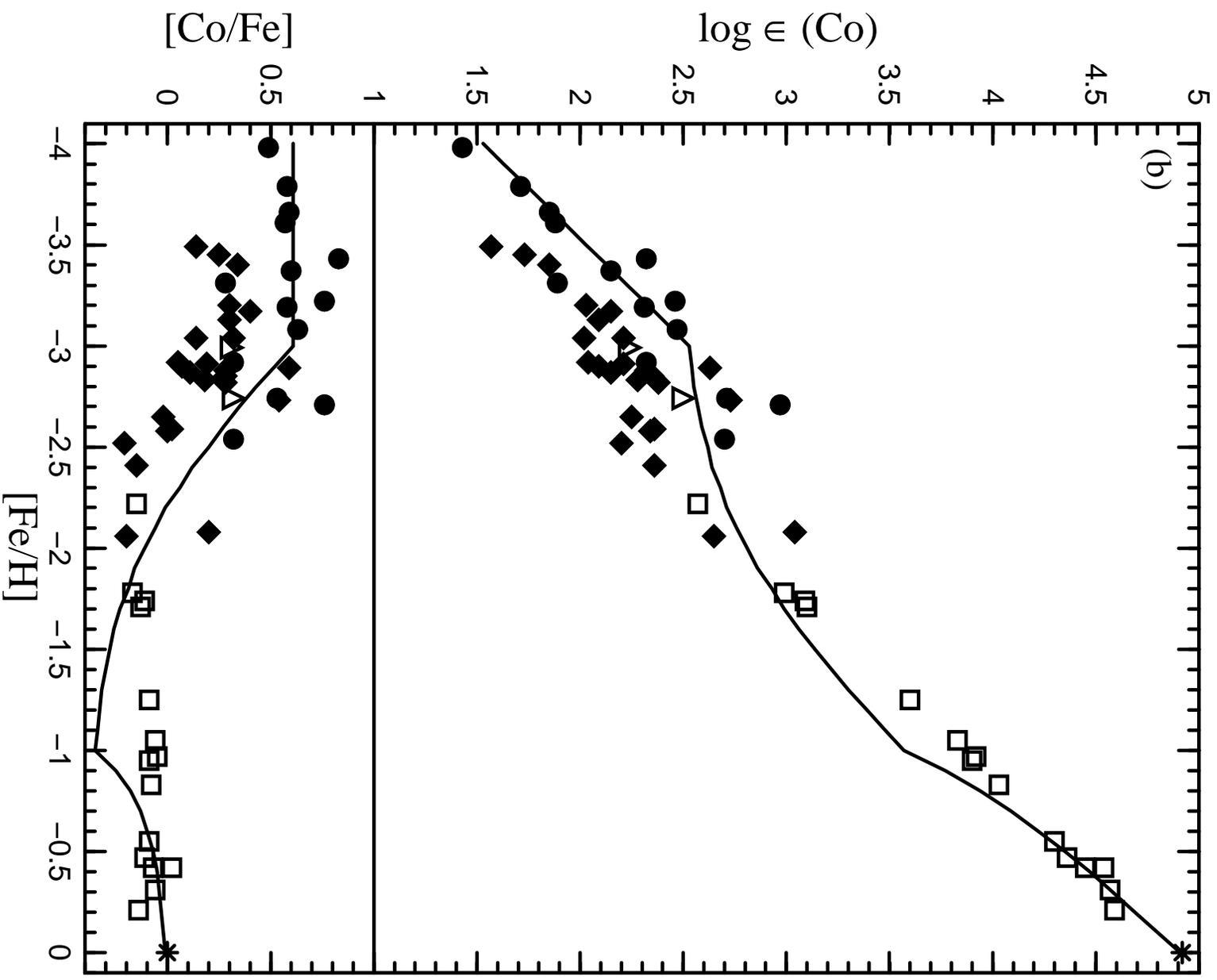
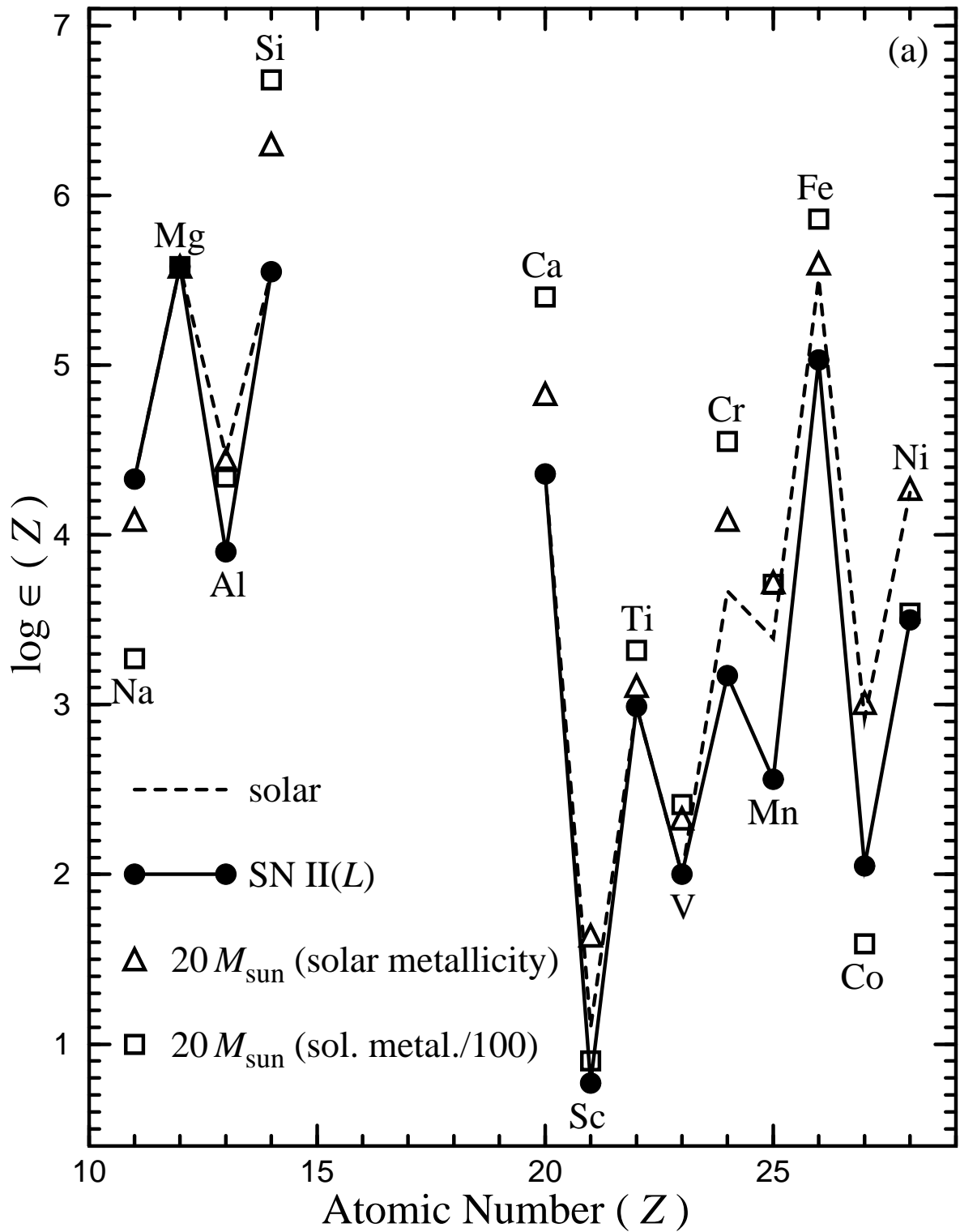
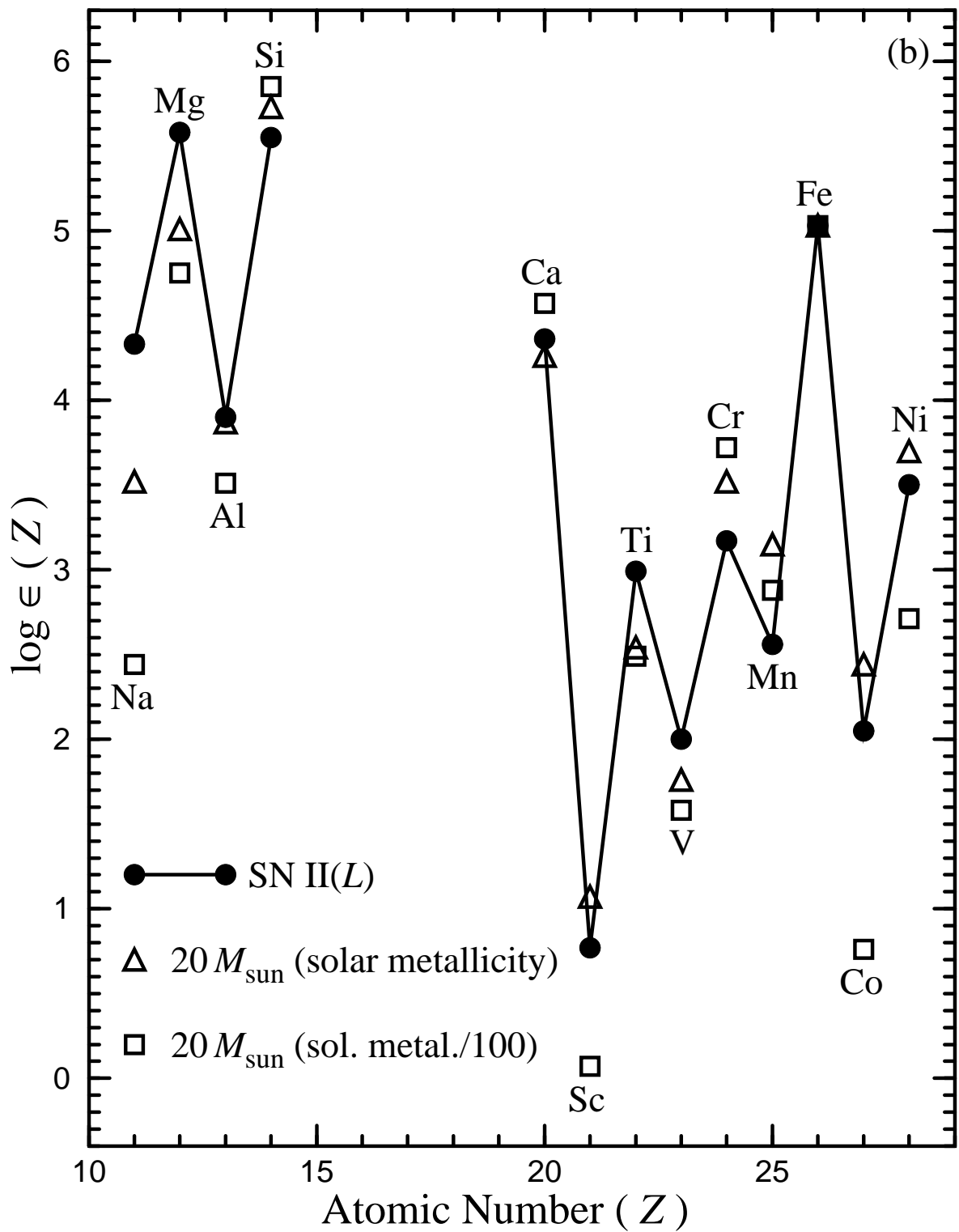
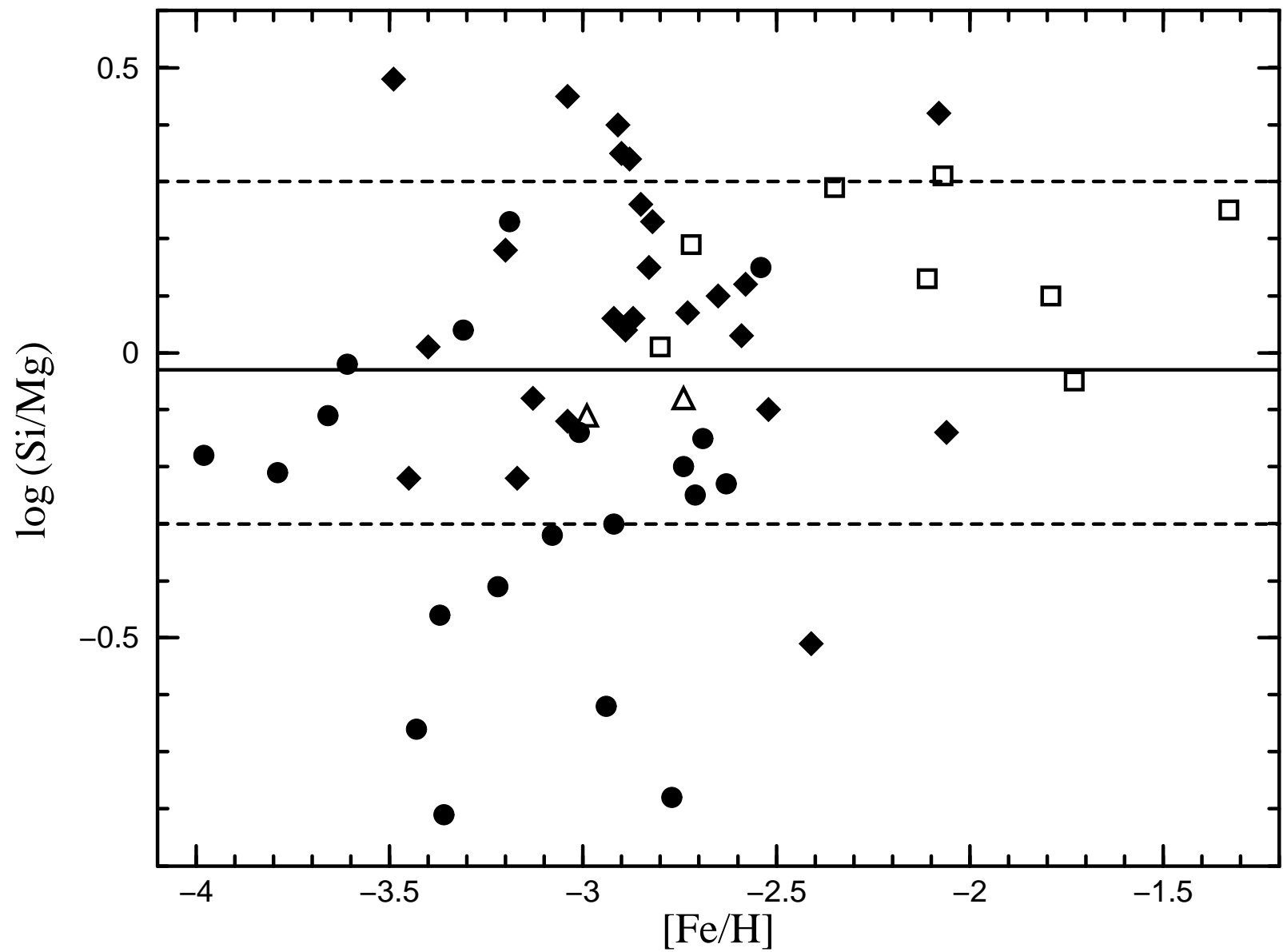
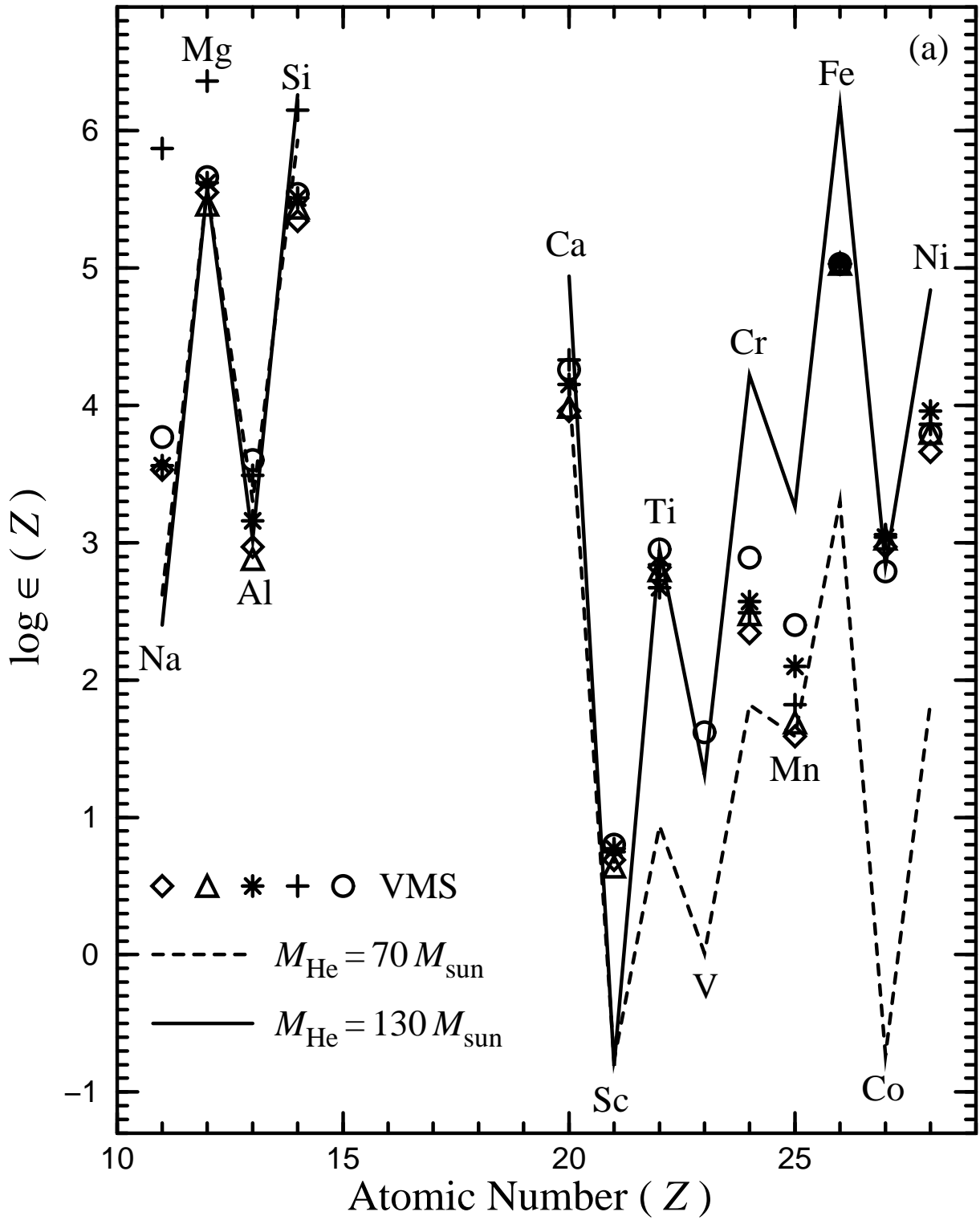
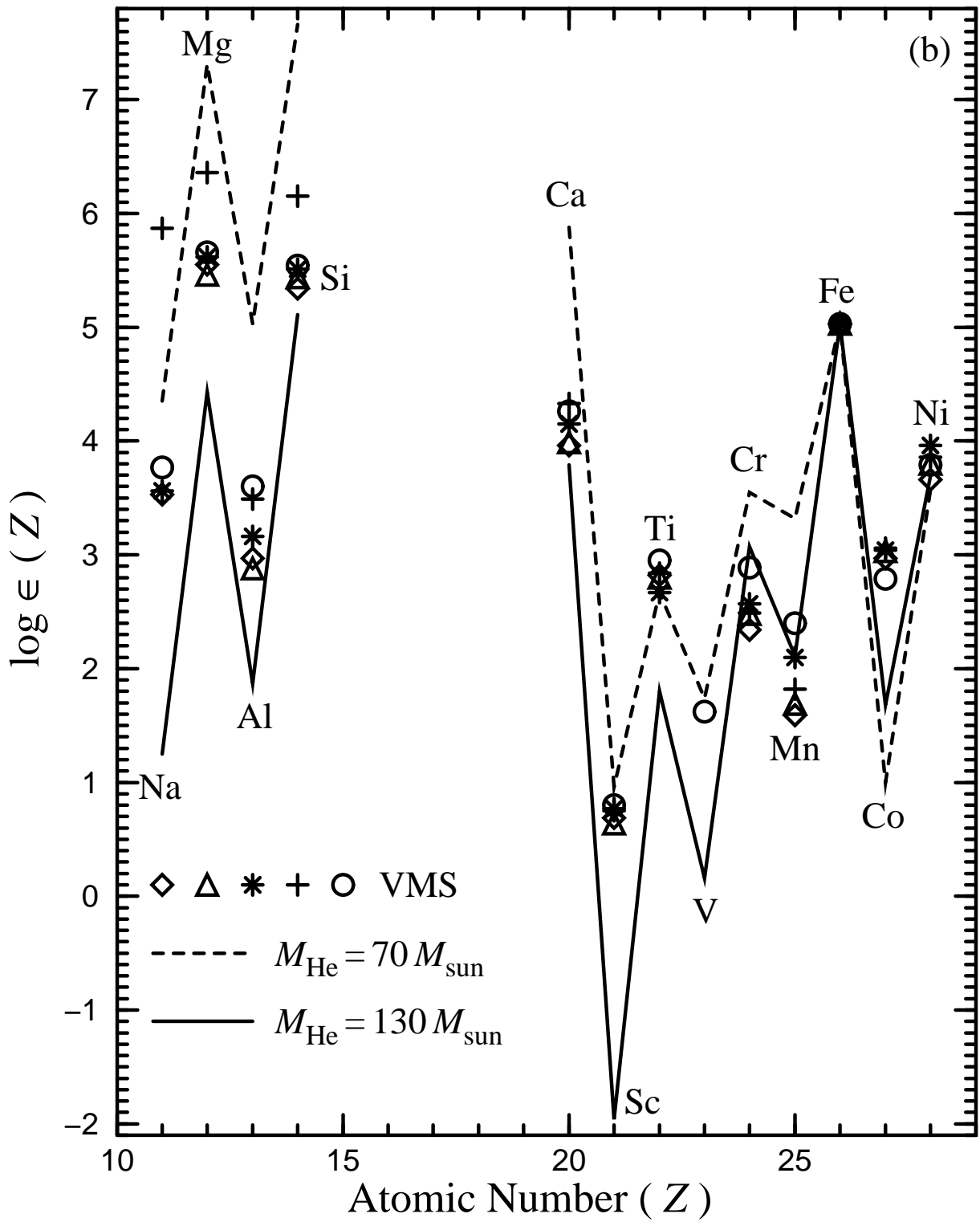
\end{document}